\documentclass[preprint, 3p]{elsarticle}
\journal{Robotics and Autonomous Systems}

\usepackage{graphicx} 
\usepackage{xcolor}
\usepackage{bm}
\usepackage{mathtools}
\usepackage{amssymb,amsfonts,amsmath}
\usepackage{units}
\usepackage{etoolbox}
\usepackage[normalem]{ulem}

\newcommand{\N}{\mathbb{N}}

\usepackage{pgfplots,pgfplotstable}
\usepgfplotslibrary{statistics}
\usetikzlibrary{arrows, plotmarks, arrows.meta,bending}
\pgfplotsset{compat=newest}
\tikzstyle{every node}=[font=\footnotesize]

\newbool{compilePlots}
\boolfalse{compilePlots} 

\begin{document}

\definecolor{K0}{RGB}{0,0,0}
\definecolor{K1}{RGB}{0,0,255}
\definecolor{K2}{RGB}{255,0,0}
\definecolor{NoisyVel}{RGB}{100,100,100}

\definecolor{XtrajRef}{RGB}{255,0,0}
\definecolor{Xtraj}{RGB}{0,0,255}
\def\lineWidthTraj{1.4}
\def\lineWidthRef{1.4}
\def\opacityRef{0.3}

\def\PlotXpoint{1} 
\def\PlotEpoint{1} 

\definecolor{Filter10}{RGB}{138,51,36}%
\definecolor{Filter30}{RGB}{253,174,97}%
\definecolor{Filter40}{RGB}{171,221,164}%
\definecolor{Filter50}{RGB}{43,131,186}%
\definecolor{Filter100}{RGB}{43,131,0}%

\definecolor{D1}{RGB}{138,51,36}%
\definecolor{D3}{RGB}{253,174,97}%
\definecolor{D13t}{RGB}{171,221,164}%
\definecolor{D5}{RGB}{43,131,186}%

\definecolor{ode}{RGB}{0,0,0}
\definecolor{Wproj}{RGB}{123,50,148}
\definecolor{WOProj}{RGB}{0,136,55}

\definecolor{quadratic}{RGB}{215,25,28}
\definecolor{nonquadratic}{RGB}{44,123,182}
\definecolor{allCostParallel}{RGB}{0,136,55}

\definecolor{multiODE}{RGB}{0,0,0}
\definecolor{multiProj}{RGB}{0,0,255}
\definecolor{multiWoProj}{RGB}{255,0,0}

\def\PlotNPointCL{10} 
\def\lwCLk{\lineWidthRef}

\def\PlotNPointCLsec{10} 

\def\MarkerRepeatKin{100} 
\def\MarkerRepeatDyn{100} 

\definecolor{Pme}{RGB}{0,0,200}
\definecolor{Pq}{RGB}{217,95,2}
\definecolor{Nme}{RGB}{139,69,19} 
\definecolor{Nq}{RGB}{231,41,138}
\definecolor{Nall}{RGB}{102,166,30}
\definecolor{Pall}{RGB}{230,171,2}

\definecolor{d5}{RGB}{166,206,227}
\definecolor{d10}{RGB}{31,120,180}
\definecolor{d50}{RGB}{178,223,138}
\definecolor{d100}{RGB}{51,160,44}
\definecolor{d200}{RGB}{251,154,153}
\definecolor{d500}{RGB}{227,26,28}
\definecolor{d1000}{RGB}{253,191,111}
\definecolor{d2000}{RGB}{255,127,0}
\definecolor{dall}{RGB}{252,15, 192}

\def\lengthPlotRef{0.5cm}

\def\markRepeatCLKinxH{24} 
\def\markRepeatCLKinvalH{10} 
\def\markDeltaTKin{1} 
\pgfmathsetmacro{\markRepeatCLKinx}{\markRepeatCLKinxH*\markDeltaTKin}
\pgfmathsetmacro{\markRepeatCLKinval}{\markRepeatCLKinvalH*\markDeltaTKin}

\pgfplotsset{
    PlotStyleKinSim/.style={
        mark repeat=\markRepeatCLKinx,
        mark options={solid, scale = 1.0, line width = \lwEDF},
        line width = \lineWidthRef,
        each nth point=\PlotNPointCL,
    }
}
\pgfplotsset{
    PlotStyleKinSimVal/.style={
        mark repeat=\markRepeatCLKinval,
        mark options={solid, scale = 1.0, line width = \lwEDF},
        line width = \lineWidthRef,
    }
}

\def\lineWidthSamples{1.0}
\def\heightMainPlot{5.5cm}
\def\heightSubplots{2.5cm}
\def\widthSubplots{5cm}

\def\deltaT{0.05}
\def\lineWidthErr{0.5}
\def\oneStepHeightD{4cm}
\pgfplotsset{
    PlotStyleOneStepErrorPosition/.style={
        line width=\lineWidthErr,
        each nth point=\PlotEpoint,
        table/x expr=\thisrowno{0}*\deltaT,
        table/y index=3
    }
}
\pgfplotsset{
    PlotStyleOneStepErrorVelocity/.style={
        PlotStyleOneStepErrorPosition,
        dotted,
    }
}
\def\YminT{-0.4} 
\def\YmaxT{0.4}
\def\XminT{0}
\def\XmaxT{2}
\def\YminTs{-0.7} 
\def\YmaxTs{1}
\def\XminTs{-0.3}
\def\XmaxTs{1.5}
\pgfplotsset{
    AxisStyleOpenLoopDi/.style={
        height= 7.0cm,       
        xmajorgrids,        
        xmin = \XminT,
        xmax = 1.4,
        ymin = \YminT,
        ymax = \YmaxT,
        axis equal image,
        ymajorgrids,
        xlabel = {$x_1$ $(\unit{m})$},
        ylabel = {$x_2$ $(\unit{m})$},
        legend cell align={left},
        legend columns = 1,
        legend style={
          fill opacity=1,
          draw opacity=1,
          text opacity=1,
          at={(1.05,0.5)},
          anchor=west,
          column sep = 0.2cm
        }
    }
}
\pgfplotsset{
    AxisStyleOpenLoopDs/.style={
        width=0.4\textwidth,   
        height= 5cm,       
        xmajorgrids,        
        xlabel = {$x_1$ $(\unit{m})$},
        ylabel = {$x_2$ $(\unit{m})$},
        xmin = \XminTs,
        xmax = \XmaxTs,
        ymin = \YminTs,
        ymax = \YmaxTs,
        axis equal image,
        ymajorgrids,
        yticklabel pos=right,
    }
}

\pgfplotsset{
    AxisStyleOneStepError/.style={
        width = 4cm,
        height = \oneStepHeightD,
        xlabel = {time (s)}, 
        ymin=1e-4, ymax=1e-1,
        xmajorgrids,
        ymajorgrids
    }
}

\def\YminE{1e-4}
\def\YmaxE{5e-1}
\def\XminE{0}
\def\XmaxE{20}
\def\facOpac{0.05}
\def\lwMulti{0.3}
\pgfplotsset{
    AxisStyleMultiStep/.style={
        width = 4.4cm,
        height = 4cm,
        xmajorgrids,
        xmin = \XminE,
        xmax = \XmaxE,
        ymin = \YminE,
        ymax = \YmaxE,
        ymajorgrids,
    }
}

\pgfplotsset{
    PlotMultiStep/.style={
        each nth point=2,
    }
}

\def\markRepeatCLDynxH{48} 
\def\markRepeatCLDynvalH{20} 
\def\markDeltaTDyn{4} 
\pgfmathsetmacro{\markRepeatCLDynx}{\markRepeatCLDynxH*\markDeltaTDyn}
\pgfmathsetmacro{\markRepeatCLDynval}{\markRepeatCLDynvalH*\markDeltaTDyn}
\pgfplotsset{
    PlotStyleDynCLsim/.style={
        line width = \lineWidthRef,
        each nth point=\PlotNPointCLsec,
        mark repeat=\markRepeatCLDynx,
        mark options={solid, scale = 0.75, line width = 0.4mm},
    }
}

\pgfplotsset{
    PlotStyleDynCLsimVal/.style={
        line width = \lineWidthRef,
        mark repeat=\markRepeatCLDynval,
        mark options={solid, scale = 0.75, line width = 0.4mm},
    }
}

\def\lwEDF{0.3mm}
\pgfplotsset{
    PlotStyleEcdf/.style={
        const plot, 
        line width = \lwEDF,
    }
}
\pgfplotsset{
    PlotStyleEcdfKin/.style={
        PlotStyleEcdf,
        mark repeat=\MarkerRepeatKin
    }
}
\pgfplotsset{
    PlotStyleEcdfDyn/.style={
        PlotStyleEcdf,
        mark repeat=\MarkerRepeatDyn
    }
}

\def\markRepeatLabKinxH{24} 
\def\markRepeatLabKinvalH{20} 
\def\markDeltaTLabKin{5} 
\pgfmathsetmacro{\markRepeatLabKinx}{\markRepeatLabKinxH*\markDeltaTLabKin}
\pgfmathsetmacro{\markRepeatLabKinval}{\markRepeatLabKinvalH*\markDeltaTLabKin}
\def\scaleMarkLabKin{0.5}
\pgfplotsset{
    PlotKinLabY/.style={
        each nth point=10,
        mark repeat=\markRepeatLabKinx,
        mark options={solid, scale=\scaleMarkLabKin, line width = \lwEDF},
    }
}
\pgfplotsset{
    PlotKinLabVal/.style={
        each nth point=1,
        mark repeat=\markRepeatLabKinval,
        mark options={solid, scale=\scaleMarkLabKin, line width = \lwEDF},
    }
}

\def\markDeltaTLabDyn{10} 
\pgfmathsetmacro{\markRepeatLabDynx}{\markRepeatLabKinxH*\markDeltaTLabDyn}
\pgfmathsetmacro{\markRepeatLabDynval}{\markRepeatLabKinvalH*\markDeltaTLabDyn}

\def\lwLabDyn{0.4mm}
\def\scaleMarkLabDyn{0.75}

\pgfplotsset{
    PlotDynLabY/.style={
        line width = \lwLabDyn,
        each nth point=1, 
        mark repeat=\markRepeatLabDynx,
        mark options={solid, scale=\scaleMarkLabDyn, line width = \lwEDF},
    }
}
\pgfplotsset{
    PlotDynLabVal/.style={
        line width = \lwLabDyn,
        mark repeat=\markRepeatLabDynval,
        mark options={solid, scale=\scaleMarkLabDyn, line width = \lwEDF},
    }
}

\begin{frontmatter}
\title{%
Data-Driven Predictive Control of Nonholonomic Robots Based on a Bilinear Koopman Realization: Data Does Not Replace Geometry} 
\author{Mario Rosenfelder\fnref{ITM}}
\author{Lea Bold\fnref{Ilmenau}}
\author{Hannes Eschmann\fnref{ITM}}
\author{Peter Eberhard\fnref{ITM}}
\author{Karl Worthmann\fnref{Ilmenau}}
\author{Henrik Ebel\fnref{LUT}\corref{correspondingauthor}}

\cortext[correspondingauthor]{Corresponding author}
\fntext[ITM]{Institute of Engineering and Computational Mechanics~(ITM), University of Stuttgart, Germany,\
\lbrack mario.rosenfelder, hannes.eschmann, peter.eberhard\rbrack @itm.uni-stuttgart.de}
\fntext[Ilmenau]{Optimization-based Control Group, Institute of Mathematics, Technische Universit\"at Ilmenau, Germany,\
\lbrack lea.bold, karl.worthmann\rbrack @tu-ilmenau.de}
\fntext[LUT]{Department of Mechanical Engineering, LUT University, Lappeenranta, Finland,\
henrik.ebel@lut.fi}

\begin{abstract}
Advances in machine learning and the growing trend towards effortless data generation in real-world systems has led to an increasing interest for data-inferred models and data-based control in robotics.
It seems appealing to govern robots solely based on data, bypassing the traditional, more elaborate pipeline of system modeling through first-principles and subsequent controller design.
One promising data-driven approach is the Extended Dynamic Mode Decomposition (EDMD) for control-affine systems, a system class which contains many vehicles and machines of immense practical importance including, e.g., typical wheeled mobile robots.
EDMD can be highly data-efficient, computationally inexpensive, can deal with nonlinear dynamics as prevalent in robotics and mechanics, 
and has a sound theoretical foundation rooted in Koopman theory.
On this background, this present paper examines how EDMD models can be integrated into predictive controllers for nonholonomic mobile robots.
In addition to the conventional kinematic mobile robot, we also cover the complete data-driven control pipeline -- from data acquisition to control design -- when the robot is not treated in terms of first-order kinematics but in a second-order manner, allowing to account for actuator dynamics. 
Using only real-world measurement data, it is shown in both simulations and hardware experiments that the surrogate models enable high-precision predictive controllers in the studied cases. 
However, the findings raise significant concerns about purely data-centric approaches that overlook the underlying geometry of nonholonomic systems, showing that, for nonholonomic systems, some geometric insight seems necessary and cannot be easily compensated for with large amounts of data. 
\end{abstract}

\begin{keyword}
Data-driven Control \sep Model Predictive Control \sep Model Learning for Control \sep Optimization and Optimal Control \sep Machine Learning for Robot Control \sep Koopman-based Control \sep Extended Dynamic Mode Decomposition \sep Nonholonomic Systems \sep Mobile Robots \sep Wheeled Robots \sep Experiments
\end{keyword}

\end{frontmatter}

\section{Introduction}
Autonomous mobile robots have become indispensable in many applications -- be it as service robots for vacuuming and cleaning, for logistics and delivery, and in security and defence. 
In many of these applications, there is nowadays a tendency to move toward the application of whole robot fleets, where the individual robots are comparatively simple and mass-produced in an economical manner, e.g., by minimizing the number of moving parts and not relying on high-precision driveline components. 
This has multiple repercussions. 
On the one hand, whereas omnidirectional and fully actuated mobile robots exist on the ground and in the air, bringing advantages regarding maneuverability and thereby simplifying control and path planning, economic and simplicity considerations have led to a prevalence of underactuated, nonholonomic mobile robots~\cite[Ch.~8]{Woernle24}. 
However, precise control of nonholonomic systems presents a challenge to this day~\cite{MurrLi07,Laval06,Jean14,Bloch15}, in particular for optimal control~\cite{HussBloc08,MullWort17,RoseEbel22}. 
Nevertheless, optimal control, such as model predictive control~(MPC), has some key advantages making it very suitable for robotics, including straight-forward controller tuning, clear definition of the control goal, and the direct consideration of constraints, which allows robots to operate safely at the border of their operating regime for maximum task-solution performance. 
This motivates the usage of MPC as this paper's control method of choice. 
On the other hand, cost-efficient, mass-produced robots that do not rely on high-precision parts may divert from a nominal, ideal behavior as it may be described by a nominal model. 
This can motivate to use a data-driven or data-augmented model for an MPC controller instead of a (purely) nominal one~\cite{EschEbel21,EschEbel23}. 
Ideally, few data points are needed to obtain a model useful for model-based closed-loop control, so that newly produced robots can be "calibrated" quickly and cost efficiently. 

As indicated  
before, model predictive control of nonholonomic vehicles is by no means simple if the control task is to ensure asymptotic stabilization of arbitrary setpoints. 
The latter can be practically useful, e.g., for a measurement robot that needs to reach certain poses with a very high accuracy to perform accurate measurements. 
It was only quite recently that it was shown that, for this task --~even for the arguably simplest nonholonomic system, i.e., the differential-drive mobile robot~-- the canonically employed quadratic cost function provably does not lead to a functional MPC controller~\cite{MullWort17}. On the contrary, tailored non-quadratic stage costs~\cite{WortMehr15,WortMehr16} or terminal conditions (either using non-quadratic terminal costs or non-convex terminal regions, see, e.g., \cite{GuHu05,GuHu06}) yield a guaranteed setpoint stabilization. The latter 
would be undesirable practically as it can introduce  
feasibility issues and control failure.
Until very recently, it was even unknown how to generally design functioning cost functions for general nonholonomic vehicles, i.e., with higher degrees of nonholonomy and with drift~\cite{RoseEbel22,EbelRose23}. 
In~\cite{RoseEbel22}, it has become clear that tailored, non-norm, mixed-exponents cost functions, which can be inferred from a homogeneous approximation of the nonholonomic system~\cite{CoroGrun20,Jean14}, are key, and~\cite{EbelRose23} has extended this to nonholonomic systems with certain kinds of actuation dynamics, i.e., if acceleration is not immediate and, thus, velocity setpoints are, in the model, not reached immediately. 
The respective stability analysis can be nicely embedded in the framework~\cite{CoroGrun20} for general nonlinear systems with null-controllable homogeneous approximations.

In the above context,  this paper provides several new contributions. 
Firstly, to the knowledge of the authors, it is the first time that purely data-driven model predictive control of a nonholonomic vehicle is performed and experimentally validated with a theory-conforming mixed-exponents cost function, without using a predefined nominal model. 
Secondly, as will be seen, we employ EDMD on a Koopman background to infer the model used for MPC. As far as the authors are aware, in this framework, this paper represents the first occasion where a second-order dynamics model of a nonholonomic mobile robot accurate enough for closed-loop MPC control is learned from real-world data. 
As will be seen, the drift present in this model has some repercussions  on data acquisition, model learning, and on control design, which are all solved within this paper. 
This paper builds upon our previous work~\cite{BoldEsch23}, in which a first-order kinematic model was considered, approximated, and its performance assessed regarding open-loop predictions only. 
Compared to the latter, novelties include studying closed-loop control in a model predictive controller with cost tailored to nonholonomic systems' sub-Riemannian geometry~\cite{Jean14} in conjunction with data-inferred models as well as data-inferred models capable of taking into account acceleration-level effects in the first place, including how to obtain and process real-world data for them. 

The paper is organized as follows. 
Section~\ref{sec:preliminaries} introduces the considered system class, model predictive control, and the EDMD approach for control-affine systems.
Section~\ref{sec:diff_robot} then introduces the kinematic and second-order dynamics differential-drive robot, as well as the specialities of controlling such a nonholonomic vehicle by means of a model predictive controller. 
In Section~\ref{sec:results_kinematic}, the data-driven models for the kinematic mobile robot, inferred from experimental measurement data, are used within predictive controllers, both in simulation, to allow large-scale comparative studies under ideal conditions, and in hardware experiments for validating whether the closed-loop controllers work as promised. 
In Section~\ref{sec:results_dynamic}, real-world data is collected for the differential-drive robot with actuator dynamics, and insight is given into post-processing the data and identifying an EDMD-based model that accounts for actuator dynamics. This model is then used within predictive controllers showing the necessity of a cost function taking into account the system's sub-Riemannian geometry arising from the nonholonomic constraints. 
Section~\ref{sec:data_efficiency} further points out how less data is needed when utilizing an EDMD-based predictive controller. 
A brief summary and outlook is given in Section~\ref{sec:summary}. 

Subsequently, for integers $n,m \in \mathbb{Z}$ with $n \leq m$, we define $\mathbb{Z}_{n:m} \coloneqq \mathbb{Z} \cap [n,m]$.
Bold variables denote a vector, or in the case of a capital bold letter, a matrix, e.g., $\bm{v}$ and $\bm{A}$, respectively. 
Furthermore, a part of a vector is denoted by specifying its component's indices as a subscript, e.g., $\bm{v}_{2:4}$ containing components 2 to 4 of vector~$\bm{v}$ in their original order.

\section{Preliminaries}\label{sec:preliminaries}
In this paper, we investigate the setpoint stabilization of nonholonomic mobile robots (possibly with drift), i.e., systems that can be described by controllable input-affine models of the form
\begin{align}\label{eq:underactuated_system}
    \dot{\bm{x}} = \bm{f}(\bm{x},\bm{u})={}\bm{g}_0 (\bm{x}) + \bm{G}(\bm{x}) \bm{u}
\end{align} 
with the control input $\bm{u}\in\mathcal{U}\subset\mathbb{R}^m$ and the state $\bm{x}\in\mathcal{X}\subseteq\mathbb{R}^n$, where $m<n$ holds such that the system is underactuated. 
Further, in addition to the control vector fields $\bm{g}_i(\bm{x})\in\mathbb{R}^n$, $i\in\mathbb{Z}_{1:m}$, which compose $\bm{G}(\bm{x})\coloneqq \begin{bmatrix} \bm{g}_1 (\bm{x}) & \dots & \bm{g}_m (\bm{x}) \end{bmatrix}\in\mathbb{R}^{n\times m}$, there may exist a drift term $\bm{g}_0(\bm{x})\in\mathbb{R}^n$. 
We assume that $\textnormal{rank}(\bm{G}(\bm{x}))=m$ for all $\bm{x}\in\mathbb{R}^n$ and that $\bm{g}_0$ and $\bm{G}(\bm{x})$ are $C^\infty$ in order to keep the presentation technically simple.
These assumptions are not restrictive for nonholonomic robotic systems and, in particular, always fulfilled in the following.

Since robotic hardware is mostly operated in a discrete fashion by utilizing piecewise constant control inputs over some sampling interval of duration $\Delta t\in\mathbb{R}_{>0}$, we consider the continuous-time system~\eqref{eq:underactuated_system} as a sampled-data system with zero-order hold on the control inputs, i.e., for some discrete-time instant~$k \in \mathbb{N}_0$ with constant control $\bm{u}(t) \equiv \hat{\bm{u}}$ on $t\in[k \, \Delta t, \ (k+1)\,\Delta t)$, it follows at $\hat{\bm{x}}=\bm{x}(k\,\Delta t)$ that
\begin{align}\label{eq:exact_discr}
    \bm{x}^+&= \bm{f}^{\Delta t} (\hat{\bm{x}}, \hat{\bm{u}})
    \coloneqq \int_{k\,\Delta t}^{(k+1)\,\Delta t}  \bm{g}_0 \left(\bm{x}(t; \hat{\bm{x}}, \bm{u}) \right) + \sum_{i=1}^m \bm{g}_i \left(\bm{x}(t; \hat{\bm{x}}, \bm{u}) \right) \hat{u}_i \, \textnormal{d}t
\end{align}
holds, where $\bm{x}(\cdot; \hat{\bm{x}}, \bm{u})$ denotes the state trajectory resulting from the initial condition $\hat{\bm{x}}$ when applying $\bm{u}(\cdot)$. 
Note that, for notational simplicity, the variables' time-dependency is often not stated explicitly in the following and we denote subsequently $\bm{x}(k)\coloneqq\hat{x}$ as well as $\bm{x}(k+1)\coloneqq \bm{x}^+$ for any $k\in\mathbb{N}_0$ in instances when the transition from one time step to the next is in the focus, irrespective of the current time (step).  
In general, there does not need to exist an explicit form of the zero-order hold discretization of the considered underactuated system~\eqref{eq:underactuated_system}, although it exists for the differential-drive mobile robot~\cite[Sec.~II]{WortMehr16}.
Moreover, the differential-drive mobile robot exemplifies that exact discretizations of the continuous-time system might not be control-affine, see, e.g.,~\cite[Sec.~II]{WortMehr16}.
However, control-affinity is retained in the limit $\Delta t \to 0$, which, coarsely speaking, makes the system approximately control affine for small sampling-time intervals since the error is decaying with $\mathcal{O}(\Delta t^2)$.

\subsection{Model Predictive Control}\label{sec:MPC}
We employ the usual MPC notation in which $\bm{x}(k\,\vert\, t)$ and $\bm{u}(k\,\vert\, t)$ are the predicted state and input trajectories, respectively, of the discretized system planned at time instant~$t$ and evaluated at time steps $k\in\mathbb{Z}_{t:(t+H)}$, where $H\in\mathbb{N}$ is the prediction horizon. 
Then, the receding horizon strategy follows by solving at each time instant $t\coloneqq k \, \Delta t$, $k\in\mathbb{N}_0$, an adequate optimal control problem~(OCP), applying the first constant control part of the optimal control input, i.e., $\bm{u}(t)\coloneqq \bm{u}^\star(t\,\vert\,t)$ for the duration  $\Delta t$ of the sampling interval, and then repeating this procedure ad infinitum.
Note that we assume w.l.o.g.\ that the underlying OCP has an optimal solution $\bm{u}^\star (\cdot\,\vert\,t)$ to avoid technical difficulties.
In particular, the OCP is given by
\begin{subequations}\label{eq:OCP}
    \begin{alignat}{3}%
            &\hspace{0pt} \underset{\bm{u}(\cdot\,\vert\, t)}{\textnormal{minimize}}
            &&\hspace{6pt}\!\sum_{k=t}^{t+H}\!\ell(\bm{x}(k\,\vert\, t),\bm{u}(k\,\vert\, t)) \label{eq:mpc_cost_generic}\\
            & \textnormal{subject to}
            &&\hspace{6pt} \bm{x} (k+1\,\vert\,t) = \tilde{\bm{f}}^{\Delta t} \left(\bm{x}(k\,\vert\,t), \bm{u} (k\,\vert\,t) \right),\quad \forall \ k \in \mathbb{Z}_{t:(t+H-1)}, \label{eq:mpc_dynamics_generic}\\
            &&&\hspace{6pt} \bm{u}(k \,\vert\, t)\in\mathcal{U},\quad \forall \ k \in \mathbb{Z}_{t:(t+H)},\\
            &&&\hspace{6pt}\bm{x}(t\,\vert\, t)=\bm{x}(t)
    \end{alignat}%
    \label{eq:mpc_optprob_generic}%
\end{subequations}%
with some continuous stage cost $\ell\! :\,\mathbb{R}^{n}\times \mathbb{R}^{m} \to \mathbb{R}_{\geq 0}$ summed up over the prediction horizon of length~$H$, and as prediction model~\eqref{eq:mpc_dynamics_generic} some (approximate) discretization $\tilde{\bm{f}}^{\Delta t} $ of any controllable nonholonomic system, possibly with drift, of nominal, continuous-time form~\eqref{eq:underactuated_system}.
Therein, the cost function~\eqref{eq:mpc_cost_generic} must, at least in parts, quantify the distance or effort it takes to steer the system (closer) to the origin, where, here, w.l.o.g., it is assumed that the coordinate system is chosen such that the goal pose lies in the origin. 
Note that the dynamics of mobile robots is generally invariant under translation and rotation such that stabilizing postures other than the origin can, due to the absence of terminal ingredients, be achieved by a coordinate transformation within the proposed OCP~\eqref{eq:OCP}. 
While the absence of a terminal constraint is in general desirable for reasons of feasibility and real-time applicability, the absence of terminal ingredients is even more convenient for nonholonomic systems since known ways to design terminal regions and costs mostly rely on a controllable system linearization around the setpoint. 
However, linearizations of nonholonomic systems are not controllable, preventing the application of typical design methods for terminal ingredients in nonlinear MPC. 
It will be seen which repercussions this will have for data-driven control using EDMD in a Koopman framework.

\subsection{Koopman Surrogate Modeling and Extended DMD}\label{sec:EDMD}

In this section, we give a short overview over how to build a data-driven surrogate model of the considered system using Koopman theory and EDMD. 
Therefore, we start with considering an autonomous nonlinear dynamical system
\begin{align}\label{eq: dyn syst}
    \dot{\bm{x}}(t) = \bm{f}(\bm{x}(t))
\end{align}
for initial condition~$\hat{\bm{x}} \in \mathbb{R}^n$, where the vector field~$\bm{f}:\mathbb{R}^{n} \rightarrow \mathbb{R}^{n}$ is locally Lipschitz-continuous. 
The Koopman operator~$\mathcal{K}^t$ of system~\eqref{eq: dyn syst} at time~$t \in \mathbb{R}_{\geq 0}$ is given by the identity 
\begin{align*}
    (\mathcal{K}^t \bm{\varphi})(\hat{\bm{x}}) = \bm{\varphi}(\bm{x}(t; \hat{\bm{x}})) 
\end{align*}
for all $(t, \hat{\bm{x}}) \in \mathbb{R}_{\geq 0} \times \mathbb{R}^{n}$ and observable functions $\bm{\varphi}: \mathbb{R}^n \rightarrow \mathbb{R}$ in a suitable function space, 
where $\bm{x}(t; \hat{\bm{x}})$ denotes the flow of System~\eqref{eq: dyn syst} at time~$t$. So instead of evaluating the observable at the flow of the system, the Koopman operator is applied to the observable function to propagate this observable function (\textit{backward} in time\footnote{The Koopman operator is the adjoint of the Perron-Frobenius operator. They are also called backward and forward (Fokker-Planck) Kolmogorov operator, respectively~\cite[Section~2]{KlusNusk18}.}). 
This propagated observable is then evaluated at the initial state. The operator~$\mathcal{K}^t$ is a linear and bounded operator but, since it acts on the infinite-dimensional observable space, it is infinite-dimensional.  
In the following, we will use EDMD to approximate the Koopman operator $\mathcal{K}|_{\mathbb{V}}$ restricted on a finite-dimensional space 
spanned by a dictionary of finitely-many observable functions. 
Therefore, let $\mathcal{D} = \{\psi_j, j \in \mathbb{Z}_{1:M}\}\subset \mathcal{C}(\mathbb{X}, \mathbb{R})$ be a dictionary of $M \in \N$ continuous observable functions.
For a compact subset of the state space~$\mathbb{X} \subset \mathbb{R}^{n}$ and a fixed time step~$\Delta t \in \mathbb{R}_{\geq 0}$, 
$d \in \N$ i.i.d.\ data points~$\bm{x}^{[1]}, \, \dots \, , \bm{x}^{[d]} \in \mathbb{X}$ are sampled and stored in the data matrix $\bm{X}=\begin{bmatrix} \bm{x}^{[1]} & \dots & \bm{x}^{[d]} \end{bmatrix}\in\mathbb{R}^{M\times d}$.
For those data points, the subsequent-in-time, propagated data points $\bm{y}^{[1]}\coloneqq\bm{x}(\Delta t; \bm{x}^{[1]}), \, \dots \, , \bm{y}^{[d]}\coloneqq\bm{x}(\Delta t; \bm{x}^{[d]})$ after time step~$\Delta t$ are collected, and analogously stored in $\bm{Y}$. 
Then, the Koopman operator $\mathcal{K}^{\Delta t}|_\mathbb{V}$ is approximated by the matrix $\bm{K}^{\Delta t} \in \mathbb{R}^{M \times M}$ 
given by 
\begin{align}\label{eq:EDMD}
    \bm{K}^{\Delta t} = \Tilde{\bm{C}}^{-1}\Tilde{\bm{A}} \qquad\textnormal{ with }\qquad \Tilde{\bm{C}} = \frac{1}{d}\bm{\Psi}_X \bm{\Psi}_X^\top, \ \Tilde{\bm{A}} = \frac{1}{d} \bm{\Psi}_X \bm{\Psi}_Y^\top,
\end{align}
where the data matrices $\bm{\Psi}_X$ and $\bm{\Psi}_Y$ are arranged as
\begin{align*}
    \bm{\Psi}_X & \coloneqq \begin{bmatrix}
         \bm{\psi}\left(\bm{x}^{[1]}\right) & \dots &  \bm{\psi}\left(\bm{x}^{[d]}\right)
    \end{bmatrix}, \\
        \bm{\Psi}_Y &\coloneqq \begin{bmatrix}
            \bm{\psi}\left(\bm{x}(\Delta t;\bm{x}^{[1]})\right) & \dots & \bm{\psi}\left(\bm{x}(\Delta t;\bm{x}^{[d]})\right)
        \end{bmatrix} 
    \end{align*}
with $\bm{\psi}(\bm{x}) = \begin{bmatrix} {\psi}_1(\bm{x}) & \dots & {\psi}_M(\bm{x}) \end{bmatrix}^\top$ being the vector of all observables in $\mathcal{D}$. So, this approximation yields a surrogate model of the Koopman operator satisfying $\bm{\psi}(\bm{x}(i\cdot \Delta t; \hat{\bm{x}})) \approx \bm{K}^{i\cdot\Delta t} \bm{\psi}(\hat{\bm{x}})$ for $i \in \mathbb{N}$. 
The first probabilistic error bounds on the estimation error were proposed in~\cite{Mezi22} and extended to also cover the projection error and i.i.d.\ sampling in~\cite{ZhanZuaz23} using finite-element techniques. 
An extension to kernel EDMD can be found in~\cite{PhilScha23a}.
Only recently, the first uniform ($L_\infty$) error bounds were derived in~\cite{KohnPhil24} depending on the fill distance using interpolation results in suitable chosen reproducible kernel Hilbert spaces.

Next, we recap the extension of EDMD to control-affine systems of the form~\eqref{eq:underactuated_system}.  
One way of dealing with the control input was proposed in~\cite{KordMezi18} (EDMDc). 
This method is particularly attractive since it provides a linear surrogate model. 
However, EDMDc has its limitations as rigorously shown in~\cite{IacoToth24}, which seem to be most-relevant for state-control couplings, i.e., if a vector field~$\bm{g}_i$ depends on the state~$\bm{x}$, see, e.g., in~\cite{BrudFu21,FolkBurd21}. 
Hence, for the applications in Sections~\ref{sec:results_kinematic} and~\ref{sec:results_dynamic}, the bilinear approach proposed in \cite{surana2016koopman,williams2016extending} is used 
since it approximately preserves the control-affine structure of system~\eqref{eq:underactuated_system}, see, e.g., \cite{OttoRowl21,StraScha24}. 
Then, the Koopman operator~$\mathcal{K}^{\Delta t}_u$ for the actuated system~\eqref{eq:underactuated_system}, now depending also on the control input $\bm{u} \in \mathcal{U}$, can be approximated for the time step~$\Delta t$ and control input~$\bm{u} \in \mathcal{U}$ by
\begin{align} \label{eq:K control}
    \bm{K}_u^{\Delta t} = \bm{K}_0^{\Delta t} + \sum_{i = 1}^{m} \lambda_i \left(\bm{K}_i^{\Delta t} - \bm{K}_0^{\Delta t}\right),
\end{align}
where the coefficients $\lambda_i$ solve the linear system $\sum\limits_{i = 1}^{m} \lambda_i \bm{u}_i = \bm{u}$, where $\bm{K}_i^{\Delta t}$ denotes the Koopman operator for time step~$\Delta t$ and fixed controls $\bm{u}_i$. 
These fixed control inputs~$\bm{u}_i$, $i\in\mathbb{Z}_{0:m}$, have to provide a basis of the control-input space. 
A canonical choice would be for example $\bm{u}_i = \bm{e}_i$ where $\bm{e}_i$ denotes the $i$th unit vector of $\mathbb{R}^{m}$. 
Notice that the collected samples~$\bm{x}^{[1]}, \dots, \bm{x}^{[d]}$ for the EDMD method~\eqref{eq:EDMD} do not need to be identical for the approximations of the $m+1$ Koopman matrices $\bm{K}_i^{\Delta t}$.  
Applying Equation~\eqref{eq:K control}, the Koopman-based lifted surrogate model can then be set up as
\begin{align}\label{eq:propagation_Koopman}
    \bm{\psi}(\bm{x}(k+1)) \approx \bm{K}_u^{\Delta t} \bm{\psi}(\bm{x}(k)).
\end{align}
We will modify the MPC algorithm by replacing the system dynamics imposed in constraint~\eqref{eq:mpc_dynamics_generic} with the Koopman-based surrogate system defined on the lifted $M$-dimensional observable space~\eqref{eq:propagation_Koopman}. 
For an extension of the above mentioned error bounds to the bilinear approximation, we refer to~\cite{NuskPeit23}. Then, Koopman-based controllers with end-to-end stability guarantees can be designed, see, e.g., \cite{StraScha23} and~\cite{BoldGrun24} for an extension to MPC without terminal conditions.

\section{Differential-Drive Mobile Robot}\label{sec:diff_robot}
Although it is the main emphasis of this paper to derive data-driven surrogate models and use them within predictive control concepts for the simplest nonholonmic mobile robot, the differential-drive robot, it is inevitable to first have a closer look at the nominal models following from first-principles.
This is mainly due to two reasons.
On the one hand, these models provide valuable insights for selecting appropriate observables for the dictionary used in EDMD and, on the other hand, they offer an understanding of the underlying geometry of nonholonomic systems, which, as will be shown, must be taken into account when designing a functioning (and theoretically sound) predictive controller also when using a data-inferred model. 
Throughout this paper, we consider two different realizations of the differential-drive mobile robot that differentiate themselves in their ways of actuation.

Firstly, the kinematic model of the mobile robot is considered, i.e., its input is some desired translational and rotational velocity which is then governed by means of underlying onboard controllers.
This is the most popular approach in (wheeled) mobile robotics due to the typically powerful drives relative to the vehicles' often small inertia such that a kinematic consideration suffices in many applications. 
Furthermore, the usually employed electric motors are controlled on velocity level because the rotation speed of the rotor shaft is easy to measure and control. 

Secondly, we consider a robot realization in which the robot's translational and rotational acceleration (proportional to propulsion force) are considered as inputs, which we refer to as a second-order dynamics mobile robot in the following.
Although this might be uncommon for wheeled mobile robots at a first glance, this realization is representative for the more general class of controllable but underactuated systems with drift of the form~\eqref{eq:underactuated_system} subject to nonholonomic constraints.
This system class comprises vehicles that, e.g., are actuated with different forms of propulsion, such as marine vehicle systems, which cannot be described reasonably by a solely kinematic model, are not as lightweight as conventional wheeled mobile robots relative to their actuation power, or, at least, have to take the dynamics of the actuators into account. 
Thus, methodologically, it stands to reason to investigate a surrogate model for this second robot realization, i.e., not assuming that the underlying motor controllers attain the desired velocity instantaneously. 
Although inertia effects are not substantial for the concrete physical hardware considered in this paper, to see if second-order dynamics data-driven modeling is feasible with the employed methodology, this second robot realization is considered within this paper to take actuator dynamics into account. 
It is worth noting that learning dynamic models is more difficult because errors on acceleration level are integrated twice and may, thus, grow quadratically with time.
Thus, subsequently, first, the first-order (kinematic) description is introduced, followed by the second-order (dynamic) one, and, finally, by a discussion of the intricacies brought about for MPC by the nonholonomic constraints in both types of model. 

\subsection{Kinematic Mobile Robot}
\begin{figure}
    \centering{
    \begin{minipage}{0.4\linewidth}
    \definecolor{vel}{RGB}{55,126,184}
\definecolor{grayNotation}{RGB}{0,0,0} 
\definecolor{stateColor}{RGB}{50,50,50} 
\definecolor{KRcol}{RGB}{255,140,0} 

\def\lwNot{0.1mm} 
\def\lwState{0.1mm} 

\begin{tikzpicture}[scale = 1.0, >=triangle 45]
    \def\diamaterOuter{3cm}
    \def\axisLength{2cm}
    \def\widthWheel{0.15cm}
    \def\radiusWheel{0.5cm}
    \def\thetaDIANA{30}
    \def\xDIANA{2cm}
    \def\yDIANA{2cm}
   
    \draw[draw opacity = 0.25, very thin] (0,0) -- (0,3);
    \draw[draw opacity = 0.25, very thin] (0,0) -- (3,0);
    \draw[->] (0,0) -- (1,0);
    \draw[->] (0,0) -- (0,1);
    \fill[black] (0,0) circle (0.05cm) node[below] {$\mathcal{K}_{\textnormal{I}}$};

    \coordinate (P) at (\xDIANA, \yDIANA);

    \draw[draw = stateColor, line width = \lwState, dashed] (P) -- ++(-\xDIANA, 0) node[left, stateColor] {$x_2$};
    \draw[draw = stateColor, line width = \lwState, dashed] (P) -- ++(0, -\yDIANA) node[below, stateColor] {$x_1$};

    \begin{scope}[xshift = \xDIANA, yshift = \yDIANA, rotate = \thetaDIANA]
        \fill[white, draw = black, fill opacity = 0.7, line width = 0.3mm] (0,0) circle (0.5*\diamaterOuter);

        \def\xShiftN{0.8cm}
        \draw[<->, draw = grayNotation, line width = \lwNot] (-1.2*\xShiftN,-0.5*\axisLength) -- ++(0, 1.0*\axisLength) node[midway, above, grayNotation, rotate = \thetaDIANA - 90] {$d$};
        
        \draw[draw = grayNotation, line width = \lwNot] (-1.35*\xShiftN,-0.5*\axisLength) -- ++(1.35*\xShiftN, 0) ;
        \draw[draw = grayNotation, line width = \lwNot] (-1.35*\xShiftN,0.5*\axisLength) -- ++(1.35*\xShiftN, 0) ;
        \draw[draw = grayNotation, line width = \lwNot] (-0.5*\radiusWheel, 0.5*\axisLength+\widthWheel) -- ++(0,0.3cm);
        \draw[draw = grayNotation, line width = \lwNot] (0.5*\radiusWheel, 0.5*\axisLength+\widthWheel) -- ++(0,0.3cm);
        \draw[<-, draw = grayNotation, line width = \lwNot] (0.5*\radiusWheel, 0.5*\axisLength+\widthWheel+0.2cm) -- ++(1.0cm,0.0cm) node[midway, above, grayNotation, rotate = \thetaDIANA, xshift = 0.2cm] {$2 r_\textnormal{w}$};
        \draw[<-, draw = grayNotation, line width = \lwNot] (-0.5*\radiusWheel, 0.5*\axisLength+\widthWheel+0.2cm) -- ++(-0.5cm,0.0cm);
        \draw[ draw = grayNotation, line width = \lwNot] (-0.5*\radiusWheel, 0.5*\axisLength+\widthWheel+0.2cm) -- ++(\radiusWheel,0.0cm);
        
        \draw[line width = 0.3mm] (0,-0.5*\axisLength) -- (0, 0.5*\axisLength);
        \filldraw[fill=black] (-0.5*\radiusWheel, -0.5*\axisLength) rectangle ++(\radiusWheel, -\widthWheel);
        \filldraw[fill=black] (-0.5*\radiusWheel, 0.5*\axisLength) rectangle ++(\radiusWheel, \widthWheel);       

        \fill[black] (0.425*\diamaterOuter,0) circle (0.05cm);
        
        \fill[black] (0,0) circle (0.075cm) node[left, yshift = 0.1cm] {P};
        \begin{scope}[opacity=.85, transparency group]
            \draw[->, KRcol, very thick] (0,0) -- (0.75,0);
            \draw[->, KRcol, very thick] (0,0) -- (0,0.75) node[right] {$\mathcal{K}_{\textnormal{R}}$};
            \fill[KRcol] (0,0) circle (0.03cm);            
        \end{scope}
        
        \draw[very thin, dashed] (0,0) -- ++(0.75*\diamaterOuter, 0);
        \begin{scope}[rotate=-\thetaDIANA]
            \draw[very thin, dashed] (0,0) -- ++(0.75*\diamaterOuter, 0);
            \def\radiusTheta{2cm}
            \draw[->, draw = stateColor, line width = \lwState] (\radiusTheta, 0) arc[start angle=0, end angle=\thetaDIANA, radius=2cm] node[midway, right, stateColor] {$\theta$};
        \end{scope}

        \draw[->, vel] (0,0) -- ++( 0.4*\diamaterOuter, 0) node[near end, xshift = -0.2cm, yshift = 0.15cm, vel] {$v, a$};
        \def\radiusOmega{0.4cm}
        \draw[->,vel] (-\radiusOmega, 0) arc[start angle=180, end angle=350, radius=\radiusOmega] node[midway, right, vel] {$\omega, \dot{\omega}$};

    \end{scope}
\end{tikzpicture}
    \end{minipage} 
    \begin{minipage}{0.4\linewidth}
        \includegraphics[width = 0.9\linewidth]{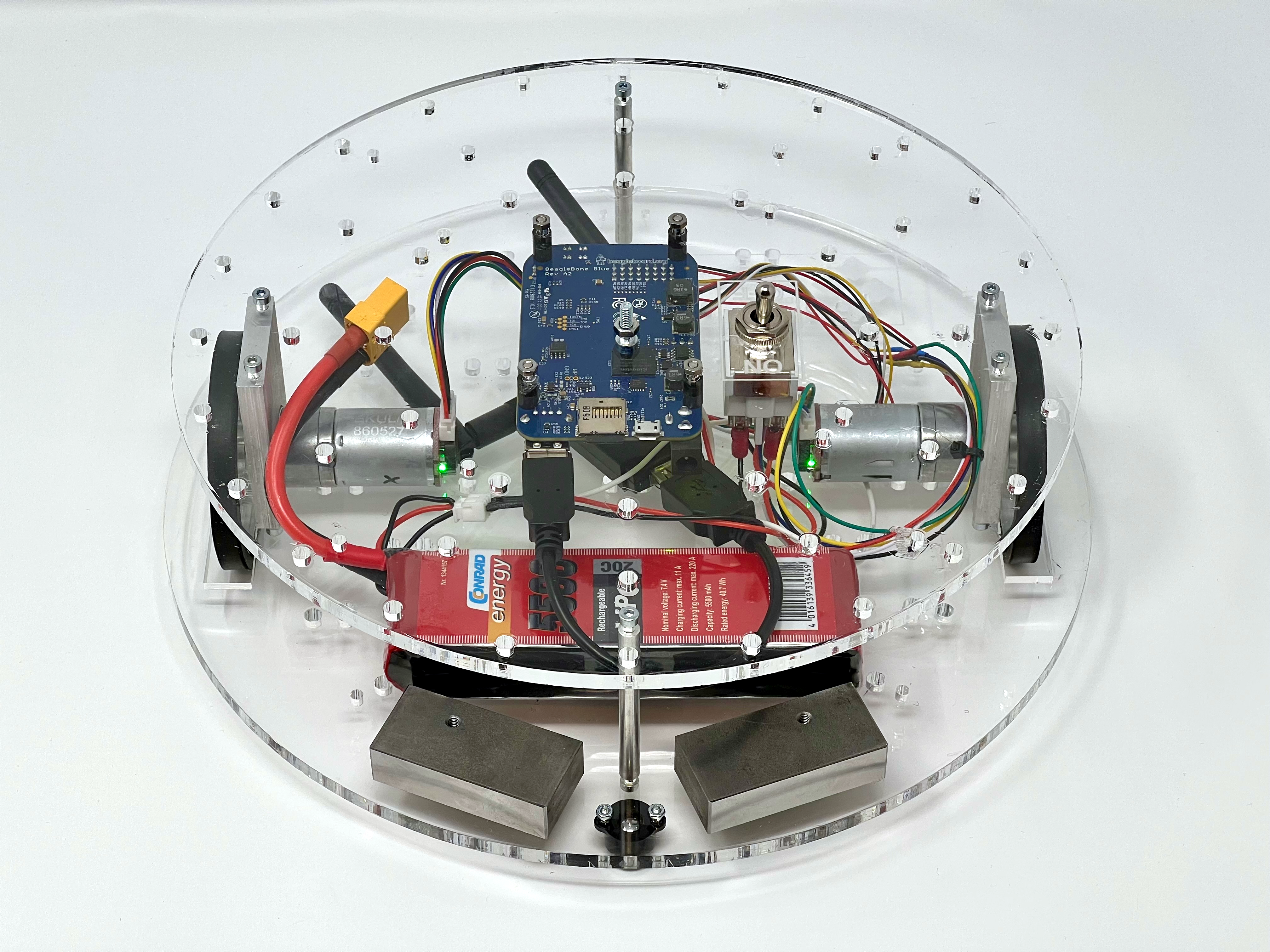}
    \end{minipage}
    }
    \caption{Schematic sketch and photograph of the employed type of custom-built differential-drive robot.}
    \label{fig:mechanical_setup}
\end{figure}
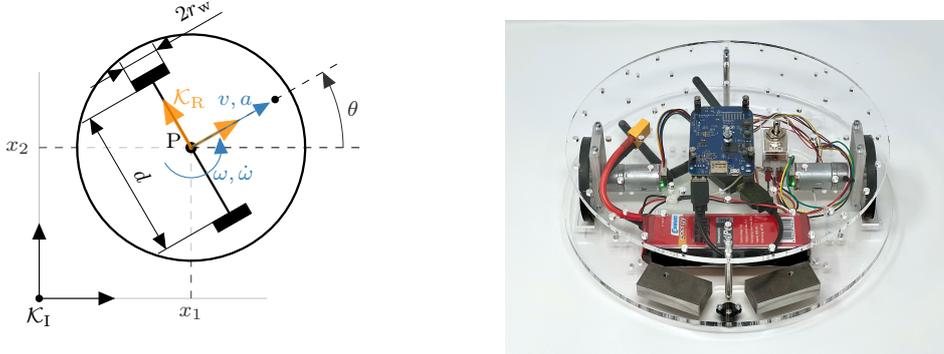

For this realization, the considered differential-drive robot is dealt with by means of its nominal kinematics describing its posture in the plane with the state $\bm{x} = \begin{bmatrix} x_1 & x_2 & \theta \end{bmatrix}^\top\in{\mathcal{X}}\subset\mathbb{R}^{n_x}$, $n_x=3$, consisting of its planar position $\begin{bmatrix} x_1 & x_2\end{bmatrix}^\top$ and its orientation $\theta$ relative to the $x_1$-axis, see Figure~\ref{fig:mechanical_setup}.
From a practical point of view, the kinematic robot is actuated by manipulating the angular velocities $\omega_i$, $i\in\lbrace \mathrm{\ell}, \, \mathrm{r}\rbrace$, of two electric motors (left and right) mounted on a common axis.
However, those angular velocities are commonly transformed into the translational velocity~$v$ and angular yaw velocity~$\omega$ of the axis' center point $P$ using the linear relation
\begin{align}\label{eq:kinematic_relation}
    \begin{bmatrix}
            v \\ \omega
    \end{bmatrix}
     = \begin{bmatrix}
        \frac{r_{\textnormal{w}}}{2} & \frac{r_{\textnormal{w}}}{2} \\[3pt]
        -\frac{r_{\textnormal{w}}}{d} & \frac{r_{\textnormal{w}}}{d} 
        \end{bmatrix}
    \begin{bmatrix}
        \omega_\mathrm{\ell} \\ \omega_{\mathrm{r}}
    \end{bmatrix},
\end{align}
where $r_{\textnormal{w}}\in\mathbb{R}_{>0}$ is the wheels' radius and $d\in\mathbb{R}_{>0}$ is the distance between the wheels, i.e., the length of the common axis, see Figure~\ref{fig:mechanical_setup}.
This yields the robot's control input $\bm{v}^\top=\begin{bmatrix} v & \omega\end{bmatrix}\in{\mathcal{V}}\subset\mathbb{R}^{m_v}$, $m_v=2$.
Due to the conventional assumption that the wheels roll without slipping, the robot is subject to the nonholonomic constraint $0=\begin{bmatrix} \sin \theta & -\cos \theta & 0\end{bmatrix} \dot{\bm{x}} \eqqcolon \bm{A}(\bm{x})\, \dot{\bm{x}}$ hindering it from moving sideways instantaneously.
\begin{subequations}\label{eq:kinematic_model}
Its nominal kinematics then follows as
\begin{align}
\dot{\bm{x}}  &=\begin{bmatrix}
    \cos (\theta)  \\ \sin (\theta)  \\ 0 
\end{bmatrix}  v +
\begin{bmatrix}
    0 \\ 0\\ 1
\end{bmatrix} \omega \\
    &\eqqcolon \bm{G}(\bm{x})\, \bm{v}, \label{eq:NH_kinematic}
\end{align} 
$\bm{x}(0) = \bm{x}_0$, where the input vector fields, i.e., the columns of $\bm{G}(\bm{x})$, are a basis for the null space of $\bm{A}(\bm{x})$.
\end{subequations}
When using such a model as a prediction model in an MPC controller, it is implicitly assumed that the robot's admissible velocities $\bm{v}\in{\mathcal{V}}$, which are in practice governed by underlying motor controllers running on a fast timescale, can be attained instantaneously. 
In practice, most differential-drive robots are modeled and controlled using the kinematic model~\eqref{eq:kinematic_model}, and deriving a precise Koopman-based surrogate model from data has been investigated in detail in our previous work~\cite{BoldEsch23}.

\subsection{Second-Order Dynamics Mobile Robot}
In this second way of approaching, modeling, and realizing control of the mobile robot, the translational acceleration~$a=\dot{v}$ in forward direction as well as the angular yaw acceleration~$\dot{\omega}$ form the designated control input $\bm{u}^\top = \begin{bmatrix} a & \dot{\omega} \end{bmatrix} \in \mathcal{U}\subset\mathbb{R}^{m_u}$, $m_u=2$.
\begin{subequations}\label{eq:2order_model}
A nominal minimal-form model of the nonholonomic robot with drift then follows using the minimal velocities~$v$ and $\omega$, see~\cite[Sec.~8.3.2]{Woernle24}, as
\begin{align}
\dot{\bm{z}}  &=
\begin{bmatrix}
    \dot{x} \\ \dot{y} \\ \dot{\theta} \\ \dot{v} \\ \dot{\omega}
\end{bmatrix} = 
\begin{bmatrix}
    v \cos (\theta) \\ v \sin (\theta) \\ \omega \\ 0 \\ 0
\end{bmatrix} + \begin{bmatrix}
    0 & 0 \\ 0& 0 \\ 0&0\\ 1&0 \\ 0&1
\end{bmatrix} \begin{bmatrix}
    a \\ \dot{\omega}
\end{bmatrix} \\
&\eqqcolon \begin{bmatrix} \bm{G}(\bm{x}) \\ \bm{0} \end{bmatrix} \bm{v} + \begin{bmatrix} \bm{0} \\ \bm{D} \end{bmatrix} \bm{u} \eqqcolon \bm{f}(\bm{z},\bm{u}), \label{eq:NH_drift}
\end{align}
$\bm{z}(0)=\bm{z}_0$.
\end{subequations}
Consequently, for the second-order dynamics model~\eqref{eq:2order_model}, the velocities of the robot are contained in its state $\bm{z}^\top = \begin{bmatrix} \bm{x}^\top & v & \omega\end{bmatrix}\in\mathcal{Z}\subset\mathbb{R}^{n_z}$, $n_z=5$.
Due to physical constraints of electric motors, e.g., regarding maximum power and angular velocity, for the class of differential-drive mobile robots, the input constraint sets~${\mathcal{V}}$ and $\mathcal{U}$ are compact, convex, and have the origin (zero input) in their respective interior. 
It is important to highlight that system~\eqref{eq:2order_model} does not represent the robot's \textit{dynamics} in the original etymological sense, i.e., describing the relationship between forces and motions.
For the special case of the planar motions of the given mobile robot with assumed center of mass at point $P$, see Figure~\ref{fig:mechanical_setup}, the inputs~$a$ and $\omega$ are proportional to the applied force and torque, respectively, and no gyroscopic effects are present.
Thus, its actual \textit{dynamics} follow by solely multiplying the fourth and fifth line of~\eqref{eq:2order_model} with the mass~$m$ and the inertia~$J$ w.r.t.\ $P$, respectively.
Due to the latter, for easy distinguishability in notation, we subsequently still refer to this robot realization as the second-order dynamics robot to clearly distinguish it from the (first-order) kinematic realization~\eqref{eq:kinematic_model}.

Although system~\eqref{eq:2order_model} is still control-affine, 
it has an additional drift term, which in practice complicates the stabilization of arbitrary postures in the plane, i.e., of setpoints of the form $\bm{z}_{\textnormal{d}}^\top=\begin{bmatrix} \bm{x}_{\textnormal{d}}^\top & 0 & 0\end{bmatrix}$.
However, the kinematic and second-order dynamics models~\eqref{eq:kinematic_model} and~\eqref{eq:2order_model} are both finite-time controllable, which can be examined via the Ra\u{s}evskij-Chow theorem~\cite{Coron07} or the Lie algebra rank condition.
In contrast, the conventional approach to determine a system's controllability via linearization fails for nonholonomic systems, see~\cite{Brockett83, RoseEbel22}, which has repercussions for MPC control design, which will be recapped from~\cite{RoseEbel22,EbelRose23} and~\cite{CoroGrun20} in the following section since it also forms the basis for the EDMD-based predictive controller considered later. 

\subsection{Sub-Riemannian MPC for Nonholonomic Vehicles}\label{sec:subRiemannian_MPC}
As hinted, when controlling nonholonomic robots using MPC as described in Section~\ref{sec:MPC}, the cost function~\eqref{eq:mpc_cost_generic} must quantify to an extent the distance or effort to steer the system to the origin.
For linear systems, or nonlinear holonomic systems that can locally be approximated by their linearization, the squared Euclidean distance is mostly taken for the state-dependent part of the cost function, which results in the canonical quadratic cost choice. 
However, the Euclidean distance does not represent a valid measure to describe the effort or distance it takes to steer nonholonomic systems such as the previously described differential-drive robot into its origin, see~\cite{MullWort17,RoseEbel22} for the corresponding proofs from a control-theoretic perspective. 
For instance, the distance a differential-drive robot has to travel to the origin is much longer in a parallel parking scenario than parking straight ahead, even in cases where the Euclidean distance might be significantly smaller in the first scenario. 

To account for this, the so-called sub-Riemannian distance~\cite[Def.~1.3]{Jean14} has to be taken into account for the class of nonholonomic systems.
This distance is given by the infimum of the length of all admissible curves on the solution manifold of the nonholonomic system which are connecting the robot's configuration and the origin while respecting the system's (nonholonomic) constraints~\cite{Jean14}.
Although the sub-Riemannian distance cannot generally and simply be computed explicitly, which is in stark contrast to the Euclidean distance and other norms, there exists a bounding pseudo-norm~\cite[Thm.~2.1]{Jean14}. 
This pseudo-norm follows from the homogeneity parameters of the system's so-called homogeneous approximation, i.e., a first-order approximation fitting to the system's sub-Riemannian geometry and thereby retaining its controllability, unlike classical linearization, which would be suitable for this task in holonomic systems and Euclidean geometry. 
As the underlying analytic calculations are not in the focus of this data-oriented paper, we kindly refer the interested reader to~\cite{Jean14,RoseEbel22,EbelRose23} for more detail about the underlying geometry. 

Based on the findings of~\cite{CoroGrun20,RoseEbel22,EbelRose23}, it can be shown that the bounding pseudo-norm yields suitable cost functions for the kinematic and second-order dynamics mobile robots~\eqref{eq:kinematic_model} and~\eqref{eq:2order_model},  reading
\begin{subequations}\label{eq:costME}
\begin{align}
    \ell_{\textnormal{me}} (\bm{x}, \bm{v}) ={} &q_1 x_1^4 + q_2 x_2^2 + q_3 \theta^4 + r_1 v^4 + r_2 \omega^4 \textnormal{\qquad and} \label{eq:costME_kinematic}\\
    \ell_{\textnormal{me}} (\bm{z}, \bm{u}) ={} &q_1 x_1^4 + q_2 x_2^2 + q_3 \theta^4 + q_4 v^4 + q_5 \omega^4 + r_1 a^4 + r_2 \dot{\omega}^4, \label{eq:costME_dynamic}
\end{align} 
\end{subequations}
respectively, where $q_i, \, r_j \in\mathbb{R}_{>0}$.
This means that, for an MPC controller utilizing OCP~\eqref{eq:OCP} subject to these costs, there exists a finite prediction horizon~$\bar{H}\in\mathbb{N}$ so that MPC controllers with~$H\geq \bar{H}$ (locally) asymptotically stabilize the robot's origin, cf.~\cite{CoroGrun20,RoseEbel22,EbelRose23}, which is not the case for the conventional quadratic cost choice~\cite{MullWort17}.
Note that here and in the following, the subscript indicates the mixed-exponents fashion of the pseudo-norm. 
It is also worth pointing out that the pseudo-norm, as the terminology suggests, does not satisfy all norm axioms on a vector space, and that, due to norm equivalence on finite-dimensional vector spaces, one cannot expect any norm to work. 

Since a usual roboticist or data scientist might not be familiar with this sophisticated geometric background, from mathematics it stands to reason to investigate how (data-inferred) predictive controllers subject to conventional quadratic cost choices perform.
The canonical choice of a robotics or control engineer, unaware of any special intricacies of nonholonomic systems, but with a-priori knowledge of a suitable set of state variables, will be a quadratic cost for penalizing the states and inputs.  
In contrast, taking a data-scientific perspective seems particularly worthwhile if no first-principles domain-expert knowledge is needed. 
In this case, the canonical choice would be, to penalize the deviations of all observables w.r.t.\ their desired setpoints with a squared Euclidean norm, as, without systemic first-principles insight, one would not know which projection of observables would constitute a (minimal) state description of the system. 
To this end, in addition to the tailored mixed-exponents costs~\eqref{eq:costME}, we consider the stage costs
\begin{subequations}\label{eq:cost_kinematic}
  \begin{align}
    \ell_{\textnormal{ce}}(\cdot) \coloneqq \ell (\bm{x},\bm{v}) ={}& \bm{x}^\top \bm{Q} \bm{x} + \bm{v}^\top \bm{R} \bm{v}, \label{eq:cost_quadr_kinematic}\\
    \ell_{\textnormal{ds}}(\cdot) \coloneqq  \ell (\bm{\psi},\bm{v}) ={}& \Delta\bm{\psi}^\top \bm{Q}_{{\psi}} \Delta\bm{\psi} + \bm{v}^\top \bm{R}_{{\psi}} \bm{v},\label{eq:costPsi_quadr_kinematic}
\end{align}  
\end{subequations}
for the kinematic mobile robot and
\begin{subequations}\label{eq:cost_dynamic}
\begin{align}
    \ell_{\textnormal{ce}}(\cdot) \coloneqq \ell (\bm{z},\bm{u}) ={}& \bm{z}^\top \bm{Q}_z \bm{z} + \bm{u}^\top \bm{R}_u \bm{u}, \label{eq:cost_quadr_dynamic}\\
    \ell_{\textnormal{ds}}(\cdot) \coloneqq  \ell (\bm{\psi},\bm{u}) ={}& \Delta\bm{\psi}^\top \bm{Q}_{z,{\psi}} \Delta\bm{\psi} + \bm{u}^\top \bm{R}_{u,{\psi}} \bm{u}, \label{eq:costPsi_quadr_dynamic}
\end{align}
\end{subequations}
for the second-order dynamics robot, respectively, where $\Delta\bm{\psi} \coloneqq \bm{\psi}-\bm{\psi}_\textnormal{d}$ since $\bm{\psi}_{\textnormal{d}}\coloneqq\bm{\psi}(\bm{x}_\textnormal{d}=\bm{0})$ is not necessarily zero and with the symmetric, positive definite weighting matrices $\bm{Q}_{(\cdot)}$ and $\bm{R}_{(\cdot)}$.
Note that in Equations~\eqref{eq:cost_kinematic} and~\eqref{eq:cost_dynamic}, $\ell_{\textnormal{ce}}$ and $\ell_{\textnormal{ds}}$ are motivated by admitting a canonical control-engineering (ce) or data-scientific (ds) perspective, respectively. 
 
Besides the stage cost design~\eqref{eq:mpc_cost_generic}, choosing an appropriate model~\eqref{eq:mpc_dynamics_generic} to predict the robot's behavior along the prediction horizon is crucial to design a functioning predictive controller without terminal ingredients. 
In the present context of using an EDMD-based surrogate model for the system's prediction, choosing the dictionary is therefore important, and discussed subsequently. 

\subsection{Choice of EDMD dictionaries}\label{sec:dictionaries}
In general, dictionaries can be chosen without incorporating detailed a-priori knowledge about the considered nonholonomic mobile robot and its underlying geometry, e.g., naively including only monomoials of all available measurements up to a predefined, usually high, order.
However, for wheeled mobile robots, and robotic systems in general, it makes sense to exploit translational invariances, periodicity of orientations, and a-priori knowledge, e.g., in terms of some nominal dynamics.
On the one hand, this can prevent overfitting and, on the other hand, this avoids unnecessarily extensive expansions of the dictionary, e.g., when the extensions would violate known invariances. 
Parsimoniously choosing a dictionary can be particularly helpful when, as in this work, the model becomes part of an OCP which is to be solved in real time. 

The prediction quality of different dictionaries has been investigated in~\cite{BoldEsch23} for the kinematic mobile robot.
Therein, dictionaries based on the right-hand side of the Euler-forward discretization, monomial dictionaries, and dictionaries based on trigonometric considerations are examined.
Due to its simplicity and since it attained very good results~\cite{BoldEsch23}, subsequently and for the kinematic mobile robot, we consider only the dictionary
\begin{align}\label{eq:D5t_kinematic}
    \mathcal{D}_{5,\textnormal{t}} = \left\lbrace 1,\, x_1, \, x_2, \, \cos(\theta), \, \sin(\theta) \right\rbrace,
\end{align}
where, here and in the following, the subscript indicates the number $M$ of observables and how the choice of dictionary is motivated (in this case from trigonometric considerations~\cite{BoldEsch23}).

Since the second-order dynamics mobile robot has not previously been investigated in this context, we consider a variety of dictionaries in the following.
In addition to a dictionary containing only the terms of the right-hand side of the Euler-forward discretization, a monomial dictionary as well as dictionaries considering more trigonometric functions are considered and defined by
\begin{subequations}\label{eq:dict_2order}
\begin{align}
    \mathcal{D}_{8,\textnormal{Eul}} = &\{  1,x_1,x_2,\theta,v,\omega, v\cos \theta, v \sin \theta \}, \label{eq:dict_Euler}\\
    \mathcal{D}_{10,\textnormal{m}} = &\{  1,x_1,x_2,\theta,v,\omega,v\omega,v\theta^2, v\theta^3 , v\theta^4 \}, \label{eq:dict_D10m}\\
    \mathcal{D}_{13,\textnormal{t}} = &\{  1,x_1,x_2,\sin\theta,\cos\theta,v,\omega,v\cos\theta,v\sin\theta , \omega\sin\theta, \omega\cos\theta, \sin\theta\cos\theta, \cos^2\theta \}, \textnormal{ and} \label{eq:dict_D13t}\\
    \mathcal{D}_{12,\textnormal{f}} = &\{  1,x_1,x_2,\theta,v,\omega,v\cos\theta,v\sin\theta, 
    v\cos(2\theta), v\sin(2\theta), v\cos(3\theta), v\sin(3\theta) \}. \label{eq:dict_D12f}
\end{align}
\end{subequations}
Dictionary~$\mathcal{D}_{10,\textnormal{m}}$ contains monomials regarding the robot's orientation (instead of trigonometric functions) up to the order of four since this yields satisfying results for the kinematic mobile robot~\cite{BoldEsch23}.
Dictionary~$\mathcal{D}_{13,\textnormal{t}}$ contains additional trigonometric functions compared to the baseline dictionary~$\mathcal{D}_{8,\textnormal{Eul}}$ in order to investigate whether this can improve the prediction accuracy.
Finally, in $\mathcal{D}_{12,\textnormal{f}}$, the dictionary is enriched in the manner of a Fourier series of the periodic variable~$\theta$, containing also trigonometric functions of higher frequencies, which is prudent as a general approximation method for periodic functions. 
Dictionary~$\mathcal{D}_{13,\textnormal{t}}$ does not contain the overall state $\bm{z}$ such that, in this case, the four-quadrant inverse tangent is utilized for reprojection, whereas the other dictionaries use the conventional coordinate projection, compare~\cite{GoorMaho23}.
Before investigating the prediction quality of the dictionaries~\eqref{eq:dict_2order} for the second-order dynamics robot, the previously unassessed closed-loop performance of the kinematic mobile robot using an EDMD-based predictive controller is looked at. 

\section{Closed-Loop Results for the Kinematic Mobile Robot}\label{sec:results_kinematic}
\begin{figure*}
    \centering
    \ifbool{compilePlots}{ 
        \tikzsetnextfilename{kinematic_sim_general}
        \include{figures/CL_first_order}
    }{
        \includegraphics{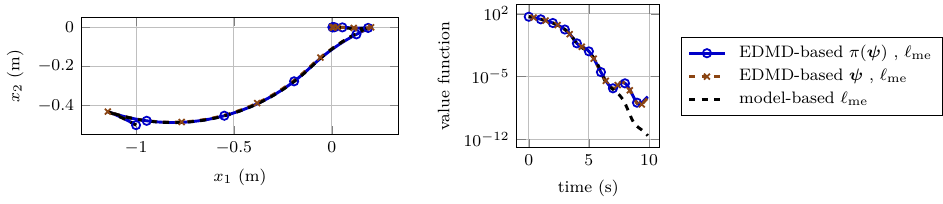}
    }
    \caption{Simulation. Closed-loop trajectories for parking the kinematic mobile robot~\eqref{eq:kinematic_model} to the origin with different predictive controllers. Left: Planar plot. Right: Value functions.}
    \label{fig:CL_1order}
\end{figure*}
Although open-loop considerations are a meaningful measure regarding the quality of a surrogate model, most of the time a mobile robot is used in a closed-loop fashion, motivating the subsequent analysis. 
In all following simulation and hardware results, the underlying OCPs are formulated using CasADi~\cite{AndGill19} and solved with IPOPT~\cite{WachBieg06}, both through the Matlab interface.

Using dictionary~\eqref{eq:D5t_kinematic}, the corresponding lifted surrogate model is derived and embedded into the predictive controller setting.
In addition to the MPC cost design for nonholonomic vehicles, a crucial choice within the EDMD-based predictive controller is the way of reprojection.
Not reprojecting in each step can significantly simplify the model propagation within the controller using~\eqref{eq:propagation_Koopman}, i.e., only based on matrix multiplications instead of numeric integration schemes.
However, the robot's orientation is explicitly needed in each prediction step within the mixed-exponents cost~\eqref{eq:costME_kinematic}.
Hence, in case of dictionary~\eqref{eq:D5t_kinematic}, which does not directly contain the angle as an observable, there is an indirect need for reprojection, which mitigates the benefits of propagating the system solely on the lifted manifold for the kinematically actuated differential-drive robot.
Reprojecting in general is not necessarily a linear operation, unlike it is the case for the coordinate projection~\cite{GoorMaho23}.
For instance, for the given dictionary~\eqref{eq:D5t_kinematic}, the, obviously nonlinear, four-quadrant inverse tangent is utilized to retrieve the orientation for usage in the cost function.

Figure~\ref{fig:CL_1order} exemplarily indicates the functionality of the resulting EDMD-based predictive controllers utilizing the mixed-exponents cost function~\eqref{eq:costME_kinematic}.
In this simulative scenario, the EDMD model has been derived using the real-world dataset from~\cite{BoldEsch23,darus-4538_2024} such that the control sampling time of the predictive controller is set to $\Delta t \coloneqq \unit[0.1]{s}$.
Furthermore, the prediction horizon is $H\coloneqq60$ and we consider, exemplarily, the initial condition $\bm{x}_0\coloneqq \begin{bmatrix} \unit[-1]{m} & \unit[-0.5]{m} & \unit[-\pi/6]{rad}\end{bmatrix}^\top$. 
Interestingly, the controllers with reprojection ($\pi(\bm{\psi})$) and without reprojection ($\bm{\psi}$) 
do not only both converge to the origin but also yield qualitatively the same results.
This may be ascribed to different facts.
Firstly, the given prediction horizon of $\unit[6]{s}$ is significantly shorter than the conducted open-loop considerations in~\cite{BoldEsch23} where reprojecting appeared to be advantageous for the prediction accuracy.
Furthermore, the cardinality of the chosen dictionary $\mathcal{D}_{5,\textnormal{t}}$ is not much larger than the dimensionality of the state space $\mathcal{X}\subset\mathbb{R}^{n_x}$, $n_x=3$, such that there are fewer observables than in very high-dimensional settings that can lead to diverging predictions. 
Last but not least, the intrinsic characteristic of a predictive controller applying only the first input sequence and then optimizing anew can mitigate accumulated deviations of later prediction time instants. 
What is crucial is that the nonholonomic mobile robot converges nicely to very close vicinity of the setpoint, i.e., indicating practical asymptotic stability of the EDMD-based closed loop~\cite{BoldGrun24}.
This can also be seen in the plot of the value function, which represents the optimal cost function~\eqref{eq:mpc_cost_generic} of the MPC's optimal control problem solved at each corresponding time instant.
For the given exemplary scenario, the remaining deviation after $\unit[10]{s}$ is less than $\unit[3\cdot10^{-2}]{mm}$ in the crucial, hard-to-control $x_2$-direction for all three controller configurations.
Note that here, and in all following simulation scenarios, the robots are simulated by means of their nominal equations of motions such that there is no plant-model mismatch for the model-based controller~(\tikz[baseline=-0.6ex] \draw[color = ode, line width=\lwCLk, dashed] (0,0) -- ++(0.5cm,0);) 
explaining its more precise behavior towards the end of the simulation when very accurate maneuvers are driven. 
In particular, the nominal kinematics is also not the robot's true dynamics, which motivates later experiments where closed-loop control is tried on actual hardware. 
Further, in Figure~\ref{fig:CL_1order} and all subsequent similar representations, the legend indicates the utilized cost function~$\ell_{\star}$ and whether it has been reprojected ($\pi(\bm{\psi})$) or not ($\bm{\psi}$).
Here, and in all subsequent comparable figures, the marks shown in the planar view correspond to those depicted in the value function plot, providing a consistent reference for time-related information, as they are evenly spaced in time.

\begin{figure}
    \centering
    \ifbool{compilePlots}{ 
        \tikzsetnextfilename{kinematic_sim_parallel}
        \include{figures/CL_first_order_parallelParking}
    }{
        \includegraphics{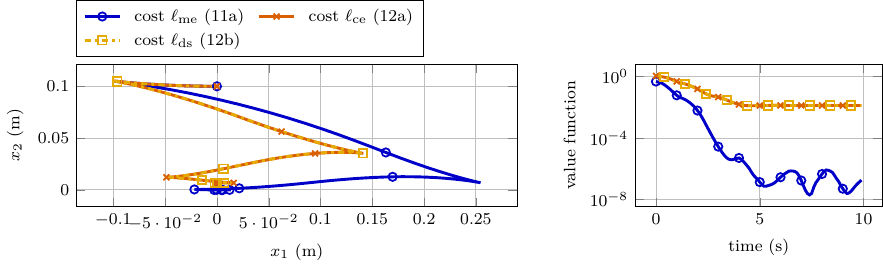}
    }
    \caption{Simulation. Left: Closed-loop trajectories for parking the kinematic mobile robot~\eqref{eq:kinematic_model} with EDMD-based predictive controllers subject to quadratic and mixed-exponents costs. Right: Corresponding value functions.}
    \label{fig:CL_1order_parallel}
\end{figure}

Since the choice of the cost function is decisive for model-based predictive control of nonholonomic systems described by first-principles models, see Section~\ref{sec:subRiemannian_MPC}, it stands to reason to investigate whether or how this affects the design of an EDMD-based predictive controller. 
Interestingly, as the simulative parallel parking example from Figure~\ref{fig:CL_1order_parallel} shows, the controllers using the conventional cost choice $\ell_{\textnormal{ce}}$~\eqref{eq:cost_quadr_kinematic} and using the abstract-minded data-scientific cost $\ell_{\textnormal{ds}}$~\eqref{eq:costPsi_quadr_kinematic} following from the lifted description, can both fail to drive the robot properly to the origin. 
From the perspective of mechanical and geometrical expertise, this may not be surprising, particularly for the classic cost choice that uses the original state variables, as there can be no escaping from the system's underlying geometry in the original state space. 
However, crucially, the quadratic metric also fails for the all-observable quadratic cost~\eqref{eq:costPsi_quadr_kinematic}, indicating that there seems to be no easy escaping from geometry even in lifted space. 

In contrast, using the tailored mixed-exponents cost~$\ell_\textnormal{me}$~\eqref{eq:costME_kinematic} evades this problem, not only for the model predictive controller based on the kinematic description following from first-principles~\cite{RoseEbel22}, but also for the EDMD-based predictive controller, see Figure~\ref{fig:CL_1order_parallel}.
The depicted results follow from the formulation with reprojection in each step, but qualitatively equivalent results are obtained without the reprojection step in the propagation of the robot's kinematics. 
Once again, it is worth pointing out that the previous results, as well as the subsequent statistical consideration, are based on an EDMD model derived from data of the real-world robot, cf.~\cite{BoldEsch23,darus-4538_2024}.

\begin{figure}
    \centering
    \ifbool{compilePlots}{ 
		\tikzsetnextfilename{kinematic_ecdf}
        \include{figures/ecdf_CL_kinematic_convergence}
    }{
        \includegraphics{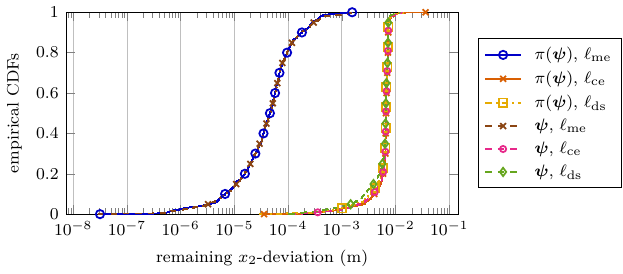}
    }
    \caption{Simulation. Empirical cumulative distribution functions (CDFs) of the deviation after $\unit[10]{s}$ in the hard-to-control $x_2$-direction when parking the kinematic mobile robot~\eqref{eq:kinematic_model} with EDMD-based predictive controllers subject to different costs and reprojection configurations.}
    \label{fig:CL_kinematic_boxplot}
\end{figure}

Importantly, the quadratic cost choices do not only fail for (strict) parallel parking cases.
To show this, 1000 random initial robot poses $\bm{x}_0\in\mathcal{X}_0\coloneqq ([-1, \,1]\,\unit{m})^2  \times [-\pi, \, \pi]\,\unit{rad}$ are drawn and each considered controller setup, i.e., regarding reprojection and cost choice, is tasked with driving the robot to the origin. 
The crucial deviation in the hard-to-control $x_2$-direction after $\unit[10]{s}$ is considered as a key measure whether the robot converged to, or at least close to, the origin. 
Figure~\ref{fig:CL_kinematic_boxplot} illustrates the results by means of the corresponding empirical cumulative distribution functions (CDFs) for the aforementioned $x_2$-deviation. 
As can be seen, all controllers subject to the quadratic costs~$\ell_{\textnormal{ce}}$ and~$\ell_{\textnormal{ds}}$ show a significant remaining deviation in this crucial direction in most cases, which is in contrast to the controllers subject to the tailored mixed-exponents cost~$\ell_\textnormal{me}$.
This emphasizes that it is not only a problem for strict parallel parking scenarios, first-principle models, and minimal state-space descriptions, and that cost design within data-driven predictive controllers can still require fundamental a-priori knowledge about the underlying geometry or mechanics of the system. 
As will be seen later for the second-order dynamics robot, this becomes even more important when controlling a nonholonomic vehicle with drift or non-negligible actuator dynamics.

\begin{figure*}
    \begin{minipage}{0.35\linewidth}
        \centering
        \includegraphics[height = 3cm, trim=300 250 300 250, clip]{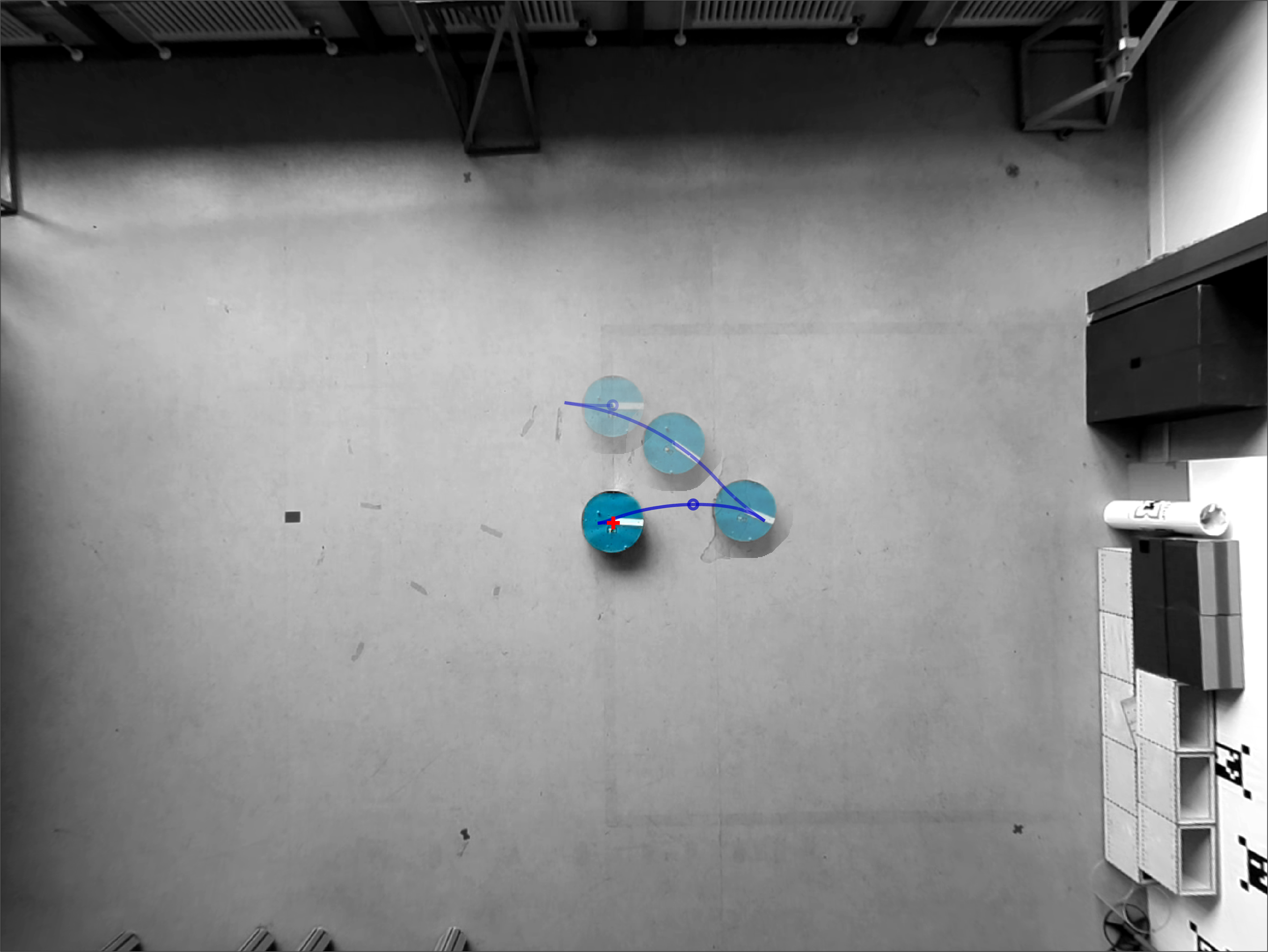}
    \end{minipage}
    \begin{minipage}{0.63\linewidth}
		\ifbool{compilePlots}{ 
    		\tikzsetnextfilename{lab_kinematic}
    		\include{figures/lab_kinematic}
    	}{
    		\includegraphics{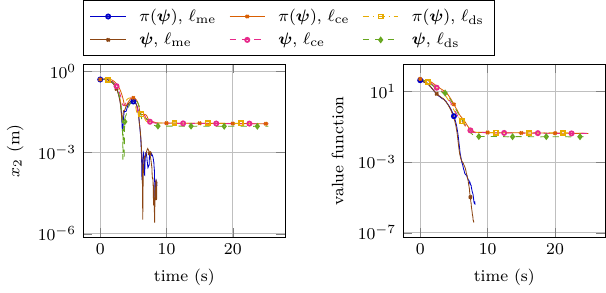}
    	}
    \end{minipage}
    \caption{Hardware results. Experimental closed-loop results for parallel parking the kinematic mobile robot with different predictive controllers. Left: Scene stills of the EDMD-based controller utilizing the mixed-exponents cost. Middle: Deviation in the hard-to-control $x_2$-direction. Right: Value function of the underlying OCPs.}
    \label{fig:kinematic_lab}
\end{figure*}
As a last consideration for the kinematic non-holonomic mobile robot, the different controllers are examined when controlling the real-world robot for an exemplary parking scenario.
To this end, first, to capture the latest behavior of the robot, an up-to-date dataset is generated using the sampling scheme described in~\cite[Sec.~5]{BoldEsch23} with a sampling time of $\Delta t \coloneqq \unit[50]{ms}$.
In order to point out a pivotal advantage of the present EDMD scheme, only five trajectories per respective input basis $\bm{v}_1=\begin{bmatrix} \unitfrac[0.2]{m}{s} & \unitfrac[-0.4]{rad}{s}\end{bmatrix}^\top$ and $\bm{v}_2=\begin{bmatrix} \unitfrac[0.2]{m}{s} & \unitfrac[0.6]{rad}{s}\end{bmatrix}^\top$ are sampled this time resulting in 150 and 642 samples, respectively.
The obtained dataset is also publicly available in~\cite{darus-4538_2024}.
Overall, the sampling lasted $\unit[119.67]{s}$ indicating the applicability of the approach for swift calibration tasks. 
The different data-based controllers are then compared for a parallel parking scenario, where the robots are initially placed by hand at $\bm{x}_0 \coloneqq\begin{bmatrix} \unit[0]{m} & \unit[0.5]{m} & \unit[0]{rad}\end{bmatrix}^\top$ within an accuracy of $\unit[1]{cm}$ and $\unit[2]{deg}$.
The corresponding experimental results are shown in Figure~\ref{fig:kinematic_lab}.
In addition to the crucial deviation in the hard-to-control $x_2$-direction and the OCPs' value function over time, the illustration on the left shows an overlay of multiple still images from the experiment, where image opacity increases as time goes by, as well as the resulting trajectory for the EDMD-based controller using the mixed-exponents cost~$\ell_{\textnormal{me}}$. 
Trajectories resulting from the other controllers are not displayed there for the sake of graphical clarity. 
Note that the experimental runs were either terminated after $\unit[25]{s}$ or once the robot reached a prescribed neighborhood of the origin, i.e., attained a predefined parking accuracy.
As Figure~\ref{fig:kinematic_lab} shows, choosing an appropriate cost function is not only decisive for artificial simulation setups but indeed plays an important role in (practical) robotics. 
It will be interesting to see subsequently how the situation will evolve when looking at second-order dynamics mobile robots in closed loop. 
\section{Open-Loop and Closed-Loop Results for the Second-Order Dynamics Mobile Robot}\label{sec:results_dynamic}
Hardware-wise, the same type of mobile robot is utilized as for the previous consideration regarding the kinematic mobile robot, see Figure~\ref{fig:mechanical_setup}. 
However, the robot's onboard software is modified for acceleration-level operation, where particularly the two independent controllers of the electric motors are adapted in order to control the motors' angular accelerations, or, more precisely, their change of angular velocity within the motor controllers' sampling time of $\unit[10]{ms}$. 
To this purpose, following from the robot's desired translational and rotational accelerations $a$ and $\dot{\omega}$, the desired translational and rotational velocity at the end of the control sampling time are computed based on relation~\eqref{eq:kinematic_relation} and governed by means of PID controllers. 
The desired accelerations are wirelessly received by the onboard program running on a BeagleBone Blue board~\cite{BeagleBoard} using the LCM library~\cite{HuangOlsonMoore10} such that the logic to acquire data (or an outer controller) can be run on an external computer to permit enough computation power without having to mount a more powerful computer onto the robot. 
Generating data for the nonholonomic second-order dynamics robot is significantly more sophisticated than for the kinematic mobile robot, mainly due to the increased dimension of the state space, the drift in the system, and since also velocity data has to be collected.
Thus, the conducted procedure to generate data for the dynamic robot is outlined in the following as it is more sophisticated than the one previously used for the kinematically described robot from our previous work~\cite{BoldEsch23}.

\subsection{Data Generation}
Starting from some arbitrary initial pose in the plane within the admissible set $\mathcal{X}_0\coloneqq[0.0, \ 1.5]\,\unit{m} \times [-0.75, \ 0.75]\,\unit{m} \times (-\pi,\pi]$ and zero velocity, a new state~$\tilde{\bm{z}}$ is drawn i.i.d.\ from the set $\mathcal{Z}_0 \coloneqq \mathcal{X}_0\times\mathcal{V}_0$, where $\mathcal{V}_0 \coloneqq [0, \ 0.4]\,\unit{m/s} \times [-1, \ 1]\,\unit{rad/s}$.
The robot then rotates, drives forward along a straight line, and again rotates towards the uniformly randomly generated point using trapezoidal desired velocity profiles, i.e., acceleration step functions.
Once the robot is sufficiently close to its desired posture $\tilde{\bm{z}}_{1:3}$, it determines an admissible acceleration to nominally reach its desired velocities $\tilde{\bm{z}}_{4:5}$ such that the actual starting point $\bm{z}_0$ to gather data is in the end different from $\tilde{\bm{z}}$.
Before applying this acceleration, it is simulated whether the corresponding input basis can be applied for some minimum number of sampling time-steps without leaving $\mathcal{Z}$ in order to obtain a significant number of data points.
If this holds, the acceleration profile is applied in an open-loop fashion to nominally reach the drawn velocities $\tilde{\bm{z}}_{4:5}$. 
When the latter are nominally reached, the actual input basis used in the EDMD model is applied, either until $\mathcal{Z}$ is left or a predefined maximum number of sampled data for that draw is exceeded.
This procedure is then repeated for some predefined number of drawn samples where the procedure alternates between the applied input bases $\bm{u}_i$, $i\in\mathbb{Z}_{0:n_u}$.
Notably, at each time step of the procedure, it is checked whether the robot is in $\mathcal{Z}$ and the current run is terminated.
If this is not the case, this causes a new state $\tilde{\bm{z}}$ to be drawn.

During the procedure, posture data is captured by an external tracking system consisting of up to five Optitrack Prime 13W cameras running at a sampling frequency of $f_\textnormal{s}=\unit[240]{Hz}$. 
This high capture frequency is utilized to collect more data along the trajectories where the input bases are applied by measuring and storing the robot's pose not only with the sampling rate of the  Koopman operator $\Delta t=\unit[50]{ms}$ but at the camera's maximum frequency. 
To that end, a separate, multi-threaded C++ program is running simultaneously in the network, solely receiving and storing the robot pose data published within the network at a sampling time of $1/f_\textnormal{s}$ as well as the commanded control inputs published at the lower frequency $1 / \Delta t$. 
Crucially, time steps of all received data are recorded with the data to be able to match-in-time states and inputs up to sampling accuracy.  
The raw data obtained this way is then processed in the following step before the surrogate model can be derived using the approach described in Section~\ref{sec:EDMD}.

\subsection{Post-Processing of Measurement Data}
\begin{figure*}
    \centering
    \ifbool{compilePlots}{ 
		  \tikzsetnextfilename{dynamic_samples}
        \include{figures/samples_2order}
    }{
        \includegraphics{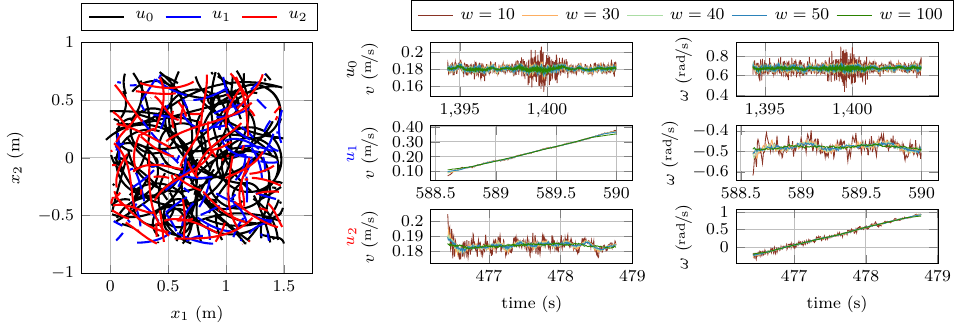}
    }
    \caption{Hardware results. Collected samples on position-level for the different input bases on the left. Smoothed translational (middle) and rotational (right) velocities for exemplary segments within the sampling procedure where the rows correspond to the input bases.}
    \label{fig:samples2order}
\end{figure*}
Since the motion capture system only provides data on position-level, the robot's  spatial velocities as well as its angular yaw velocity are approximated offline from the posture data using central differences of first-order, e.g., for time instant $t_k$, we obtain the approximate velocity in the $x_1$-direction
\begin{align}\label{eq:central_difference}
    \dot{x}_1(t_k) = \frac{x_1 (t_{k+1}) - x_1 (t_{k-1})}{t_{k+1} - t_{k-1}},
\end{align}
where the denominator is nominally equal to $2 / f_{\textnormal{s}}$.
The obtained spatial velocities described in the inertial frame of reference~$\mathcal{K}_\textnormal{I}$ are then transformed into the robot's frame of reference $\mathcal{K}_\textnormal{R}$, see Figure~\ref{fig:mechanical_setup}, using its current orientation~$\theta (t_k)$.
In the next step, the obtained velocities are smoothed using a moving average, where the size of the averaging window is a hyperparameter with its influence to be inspected. 
It is worth noting that the fact that the (desired) velocity profiles are either constant or linear for the applied input bases during data generation makes smoothing simpler since at least the desired motions do not contain extreme temporal detail that would be quickly lost for larger window sizes. 
Also, individual segments of constant or linear desired velocity are smoothed individually, to not negatively influence accuracy at the starts and ends of respective sections.

Finally, for each input~$\bm{u}_i$, $i\in\mathbb{Z}_{0:n_u}$, the corresponding (state) measurements are stored as data pairs of states and successor states, where the latter are recorded $\Delta t$ later than the former, in the matrices $\bm{X}_i$ and $\bm{Y}_i$, respectively.  
In addition to numerically approximating the robot's velocities, the measurements of the robot's orientation are post-processed, too.
Since the motion capture system provides measurements within the interval $(-\pi, \, \pi]$, there might be undesired jumps w.r.t. the orientation and its successor value.
Therefore, all angular measurements are continued continuously  and then, the entries of $\bm{X}_i$ are shifted back to their equivalent value within $(-\pi, \, \pi]$ and the corresponding entries of $\bm{Y}_i$ are shifted by the same amount such that some orientations in $\bm{Y}_i$ may lie outside of $(-\pi, \, \pi]$.
However, before evaluating the model subsequently, the orientation is shifted to $(-\pi, \, \pi]$.

For the described sampling and post-processing scheme, we have collected 108106, 13705, and 24910 samples for the chosen input bases 
\begin{align}
    \bm{u}_0 &= \begin{bmatrix}  \unitfrac[0]{m}{s^2}& \unitfrac[0]{rad}{s^2} \end{bmatrix}^\top, \quad
    \bm{u}_1 = \begin{bmatrix}  \unitfrac[0.2]{m}{s^2}& \unitfrac[0]{rad}{s^2} \end{bmatrix}^\top, 
    \quad \text{and} \quad
    \bm{u}_2 = \begin{bmatrix}  \unitfrac[0]{m}{s^2}& \unitfrac[0.5]{rad}{s^2} \end{bmatrix}^\top,
\end{align}
respectively, where for each basis 100 trajectory segments are run, see Figure~\ref{fig:samples2order}.
For the zero-input~$\bm{u}_0$, the second-order dynamics robot drives in a circular motion~(\tikz[baseline=-0.6ex] \draw[color = K0, line width=\lineWidthSamples] (0,0) -- ++(\lengthPlotRef,0);) 
with the radius being determined by the quotient of the constant translational and angular velocities, becoming a straight line in the limit case of zero angular velocity. 
Input basis~$\bm{u}_1$ linearly increases the robot's translational velocity and thus enlarges the driven radius~(\tikz[baseline=-0.6ex] \draw[color = K1, line width=\lineWidthSamples] (0,0) -- ++(\lengthPlotRef,0);) 
over time whereas $\bm{u}_2$ increases the angular velocity $\omega$ such that an S-shaped profile or a circular motion with shrinking radius follows~(\tikz[baseline=-0.6ex] \draw[color = K2, line width=\lineWidthSamples] (0,0) -- ++(\lengthPlotRef,0);), 
cf. Figure~\ref{fig:samples2order}.
For the input bases~$\bm{u}_1$ and $\bm{u}_2$, the length of a sampled trajectory is for most cases limited by the velocity constraint set~$\mathcal{V}$ such that significantly less samples are collected in contrast to the input basis~$\bm{u}_0$.
In addition, Figure~\ref{fig:samples2order} depicts the smoothed translational and rotational velocities for different window sizes $w$ of the moving average exemplarily for one sampled trajectory per input basis.
As shown there, especially the measured angular velocity $\omega$ is noise sensitive and for smooth velocity profiles, window sizes $w\geq 30$ should be considered, which correspond to the average taken over a window of $\geq \unit[125]{ms}$.
Note that the window size choice has a strong influence on the prediction quality obtained later since it directly determines the integrator part of the dynamics~\eqref{eq:2order_model}, i.e., minor variations in the integrator will inevitably lead to widely varying open-loop predictions regarding the robot's position as errors may sum up over time.  

\subsection{Open-Loop Results}\label{sub:OL_2order}
\begin{figure*}
    \centering
    \ifbool{compilePlots}{ 
		\tikzsetnextfilename{dynamic_dictionaries}
        \include{figures/dictionaries}
    }{
        \includegraphics{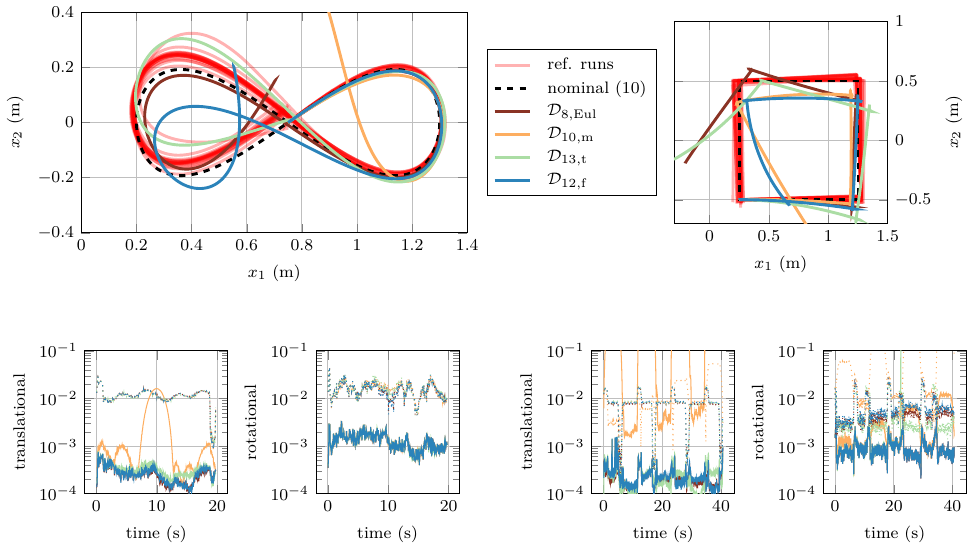}
    }
    \caption{Simulation. Top: Open-loop predictions for surrogate models subject to different dictionaries~\eqref{eq:dict_Euler} and~\eqref{eq:dict_2order}, for a $\infty$-shaped reference trajectory on the left and a square-shaped reference trajectory on the right. 
    Bottom: Corresponding translational and rotational one-step errors for the different dictionaries averaged over the 20 reference runs, where the solid lines correspond to the errors on position-level and the dotted lines are the one-step errors on velocity-level.}
    \label{fig:dictionaries}
\end{figure*}
Having indicated the sensitivity of the open-loop predictions w.r.t.\ the velocity filtering, a remaining core issue for learning the second-order dynamics robot is the dictionary choice.
Therefore, we fix the window size for velocity filtering in the following to $w\coloneqq40$ and investigate the open-loop predictions for the different dictionaries proposed in Section~\ref{sec:dictionaries}. 
Figure~\ref{fig:dictionaries} shows the resulting open-loop predictions (top plots) as well as the respective one-step errors (bottom plots) averaged over 20 runs for two reference trajectories, a $\infty$-shaped trajectory (left plots) and a square-shaped trajectory (right plots), respectively.
The experimental reference runs of the hardware robot are overlaid in red with reduced opacity~(\tikz[baseline=-0.6ex] \draw[color = XtrajRef, line width=\lineWidthRef, opacity = \opacityRef] (0,0) -- ++(\lengthPlotRef,0);) 
and, as a reference, the black, dashed trajectory~(\tikz[baseline=-0.6ex] \draw[color = black, line width=\lineWidthRef, dashed] (0,0) -- ++(\lengthPlotRef,0);) 
shows the nominal trajectory following from the time integration of the nominal equations of motion~\eqref{eq:2order_model}.
In the one-step error plots, the translational as well as the rotational one-step errors are depicted on position level (solid lines) and velocity level (dotted lines), respectively, averaged over the 20 reference runs.

Unlike the surrogate model of the kinematic robot~\cite{BoldEsch23}, monomials for the orientation up to order four appear not to be sufficient to predict the open-loop behavior of the second-order dynamics robot, see the one-step errors for dictionary~$\mathcal{D}_{10,\textnormal{m}}$~(\tikz[baseline=-0.6ex] \draw[color = D3, line width=\lineWidthTraj] (0,0) -- ++(\lengthPlotRef,0);).
In contrast, the one-step errors for the three dictionaries $\mathcal{D}_{8,\textnormal{Eul}}$, $\mathcal{D}_{13,\textnormal{t}}$, and $\mathcal{D}_{12,\textnormal{f}}$ are all in a similar order of magnitude and generally better than for~$\mathcal{D}_{10,\textnormal{m}}$.
Again, it is important to note that solely considering the open-loop prediction over a long time span can easily mislead since prediction errors can, by pure chance, cancel out over multiple time steps.
Note that the generally weaker performance for the square-shaped trajectory regarding the open-loop predictions can, on the one hand, be reasoned by the extended period of time.
On the other hand, since the robot only turns counter-clockwise within this reference trajectory, (one-step) errors in the prediction of the crucial robot's orientation and rotational velocity do not cancel out over time, leading to predictions that are increasingly superfluously rotated with increasing time. 

Including trigonometric functions with higher frequencies in dictionary~$\mathcal{D}_{12,\textnormal{f}}$~~(\tikz[baseline=-0.6ex] \draw[color = D5, line width=\lineWidthTraj] (0,0) -- ++(\lengthPlotRef,0);) 
does not yield improvements compared to the baseline dictionary~$\mathcal{D}_{8,\textnormal{Eul}}$~~(\tikz[baseline=-0.6ex] \draw[color = D1, line width=\lineWidthTraj] (0,0) -- ++(\lengthPlotRef,0);).
Thus, and since $\mathcal{D}_{8,\textnormal{Eul}}$ is very simple and shows especially good predictions within the first-half of the $\infty$-shaped reference trajectory, the following investigations are conducted with dictionary~$\mathcal{D}_{8,\textnormal{Eul}}$ and a fixed smoothing window of $w=40$. 
Although the achieved open-loop predictions for the second-order dynamics robot are not as precise as for the kinematic one~\cite{BoldEsch23}, the results seem good enough to predict the robot's dynamical behavior well for multiple seconds, which provides promise for usage within a predictive controller~\cite{BoldGrun24}, as that needs to predict the behavior only for a limited time horizon. 
This can be seen when the prediction errors are analyzed over multiple steps (of variable number), and not just from the initial state, but from all states along the trajectory.
Interested readers may find a similar multi-step error study for the kinematic mobile robot in~\cite{BoldEsch23}. 
\begin{figure*}
    \centering
    \ifbool{compilePlots}{ 
		\tikzsetnextfilename{multistep}
        \include{figures/multistep}
    }{
        \includegraphics{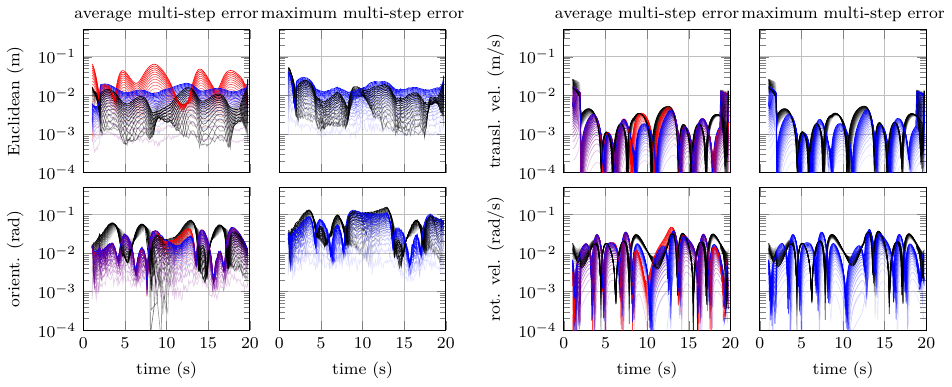}
    }
    \caption{Simulation. Averaged and maximum multi-step prediction errors of the surrogate model using dictionary~$\mathcal{D}_{8,\textnormal{Eul}}$ with (blue line) and without (red line) reprojection as well as of the nominal model~\eqref{eq:2order_model} (black line) with respect to the 20 reference runs of the $\infty$-shaped trajectory.}
    \label{fig:multistep}
\end{figure*}
Figure~\ref{fig:multistep} displays these multi-step errors for the $\infty$-shaped reference trajectories where the opacity increases with the number of prediction steps, i.e., the line with the lowest opacity marks the one-step error whereas full opacity corresponds to the error at the end of the considered prediction horizon (here $H=20$).
Note that Figure~\ref{fig:multistep} displays the multi-step errors averaged over the 20 reference runs as well as the maximum multi-step error out of these runs.
Moreover, in addition to the prediction with reprojection in each step, cf.~(\tikz[baseline=-0.6ex] \draw[color = multiProj, line width=2*\lwMulti] (0,0) -- ++(\lengthPlotRef,0);), 
the multi-step errors for the prediction without reprojection in each step, cf.~(\tikz[baseline=-0.6ex] \draw[color = multiWoProj, line width=2*\lwMulti] (0,0) -- ++(\lengthPlotRef,0);), 
as well as the errors achieved with the nominal model~(\tikz[baseline=-0.6ex] \draw[color = multiODE, line width=2*\lwMulti] (0,0) -- ++(\lengthPlotRef,0);) 
are also displayed in Figure~\ref{fig:multistep}.
It turns out that the prediction with reprojection in each step is most of the time slightly advantageous for the velocity and orientation descriptions compared to the nominal model. 
At the same time, for an increasing prediction time, the nominal model's accuracy is higher for the robot's predicted position.
Predicting without a reprojection in each step results in a decisive loss of accuracy for the second-order dynamics mobile robot, increasing the position error by almost one order of magnitude.
Thus, this case is omitted in the plots for the maximum multi-step error for reasons of graphical clarity.
For the surrogate model with reprojection~(\tikz[baseline=-0.6ex] \draw[color = multiProj, line width=2*\lwMulti] (0,0) -- ++(\lengthPlotRef,0);), 
the maximum errors at the end of the prediction horizon ($\unit[1]{s}$) for the position and the orientation are smaller than $\unit[5]{cm}$ and $\unit[8]{deg}$, respectively.
This nurtures the hope to be able to precisely control also the robot with actuator dynamics within a predictive control concept, which is examined subsequently.

\subsection{Closed-Loop Results}\label{sub:CL_dynamic}
Based on the previous findings, the following closed-loop considerations are conducted for the baseline dictionary $\mathcal{D}_{8,\textnormal{Eul}}$~\eqref{eq:dict_Euler}.
Note that the latter eases reprojecting if a conventional coordinate projection~\cite{GoorMaho23} is conducted since $\mathcal{D}_{8,\textnormal{Eul}}$ explicitly contains the overall state $\bm{z}$, in contrast to the kinematically actuated robot with dictionary~$\mathcal{D}_{5,\textnormal{t}}$ from~\eqref{eq:D5t_kinematic}. 
Besides showing that the learned surrogate model can be utilized to precisely control also second-order dynamics mobile robots, it will be seen that choosing an appropriate cost function within the predictive controller becomes even more important for nonholonomic systems with drift.
Again the conventional quadratic cost penalizing the states~\eqref{eq:cost_quadr_dynamic}, the quadratic cost~\eqref{eq:costPsi_quadr_dynamic} w.r.t.\ the lifted description, and the tailored mixed-exponents cost~\eqref{eq:costME_dynamic} are considered.

\begin{figure} \centering
    \ifbool{compilePlots}{ 
		\tikzsetnextfilename{dynamic_sim}
        \include{figures/CL_second_order}
    }{
        \includegraphics{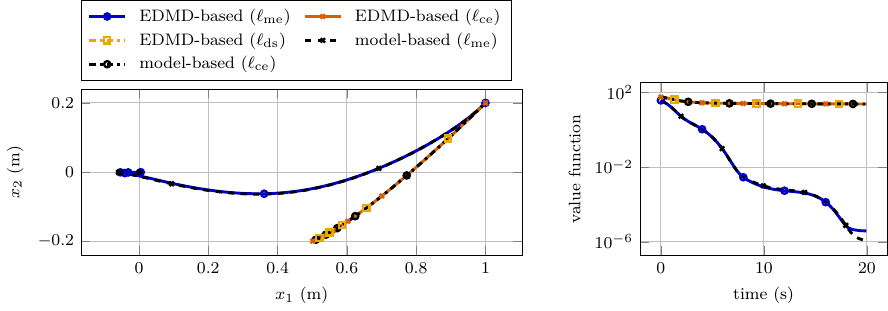}
    }
    \caption{Simulation. Closed-loop trajectories for parking the second-order dynamics mobile robot~\eqref{eq:2order_model} to the origin with different predictive controllers. Left: Planar plot. Right: Corresponding value functions.}
    \label{fig:CL_2order}
\end{figure}

Figure~\ref{fig:CL_2order} exemplarily shows the performance of the predictive controllers subject to different cost functions.
In this simulative scenario and all other following closed-loop scenarios, the EDMD model has been derived using the same real-world dataset~\cite{darus-4538_2024} as for the previous open-loop considerations.
Starting from an arbitrary initial configuration of $\bm{z}_0 \coloneqq \begin{bmatrix} \unit[1]{m} & \unit[0.2]{m} & \unit[\pi/4]{rad} & \unitfrac[0]{m}{s}  & \unitfrac[0]{rad}{s}\end{bmatrix}^\top$, the robot shall park to the origin using the respective predictive controllers with a horizon $H \coloneqq 50$, a control sampling time of $\Delta t \coloneqq \unit[50]{ms}$, and reprojecting within each prediction step.
As before, the EDMD-based predictive controllers subject to the quadratic costs~$\ell_{\textnormal{ce}}$ and $\ell_{\textnormal{ds}}$ are not capable of driving the robot to its origin.
This intrinsic problem of suitably describing the distance the robot has to travel to the setpoint can moreover not be solved by simply enlarging the prediction horizon~\cite{MullWort17,RoseEbel22} and does also occur for a model predictive controller following from first-principles, i.e., without plant-model mismatches in simulations.
Crucially, the controllers subject to quadratic costs even perform significantly worse here than previously in the kinematic case, compare Figures~\ref{fig:CL_1order_parallel} and~\ref{fig:CL_2order}.

\begin{figure}
    \centering
    \ifbool{compilePlots}{ 
		\tikzsetnextfilename{dynamic_ecdf}
        \include{figures/ecdf_CL_dynamic_convergence}
    }{
        \includegraphics{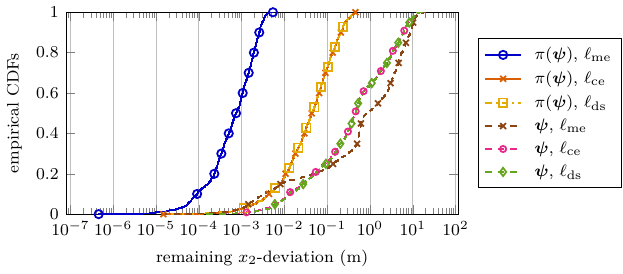}
    }
    \caption{Simulation. Empirical cumulative distribution functions of the deviation after $\unit[20]{s}$ in the hard-to-control $x_2$-direction when parking the second-order dynamics mobile robot~\eqref{eq:2order_model} with EDMD-based predictive controllers subject to different costs and reprojection configurations.}
    \label{fig:CL_2order_boxplot}
\end{figure}

This impression is further corroborated by the subsequent consideration in which the second-order dynamics robot shall park to the origin from 1000 random initial configurations $\bm{z}_0 = \begin{bmatrix} \bm{x}_0^\top & \unitfrac[0]{m}{s} & \unitfrac[0]{rad}{s}\end{bmatrix}^\top\in\mathcal{Z}_0$, $\bm{x}_0 \in\mathcal{X}_0$. 
For each drawn initial condition, the scenario is simulated using different control setups (w.r.t.\ projection and cost choice) and then, the deviation in the crucial hard-to-control $x_2$-direction after $\unit[20]{s}$ is considered.
Figure~\ref{fig:CL_2order_boxplot} shows the statistics of the different controllers by means of empirical cumulative distribution functions with respect to this remaining deviation.
Only the EDMD-based predictive controller using the tailored mixed-exponents cost and the reprojection in each time step reliably drives the mobile robot~\eqref{eq:2order_model} to the origin.
In 75\% of the cases the remaining deviation in the $x_2$-direction is less than $\unit[2]{mm}$.
For all setups subject to a quadratic cost, i.e., $\ell_\textnormal{ce}$ and $\ell_{\textnormal{ds}}$, the predictive controller often fails to drive the robot to the origin.
Moreover, the remaining deviations in the hard-to-control direction are significantly larger than it is the case for the kinematically modeled mobile robot (Figure~\ref{fig:CL_kinematic_boxplot}), i.e., empirically, choosing a tailored stage cost taking into account the sub-Riemannian geometry of nonholonomic systems becomes even more important for vehicles with drift, also when using a data-driven predictive controller.
Furthermore, for the second-order dynamics mobile robot, EDMD-based predictive controllers without reprojection along the prediction horizon are not able to reliably drive the robot to the origin, see Figure~\ref{fig:CL_2order_boxplot}, which is in contrast to the kinematically modeled mobile robot.
A possible reason for the difference could be the larger number of observables that are not states, due to the increased dimension of the lifted space compared to the state space dimension. 
However, the EDMD-based predictive controller without reprojection in each step appears to be very appealing from a practical point of view such that further investigations are the topic of ongoing work.

\begin{figure*}
    \begin{minipage}{0.35\linewidth}
        \centering
        \includegraphics[height = 3cm, trim=175 200 400 325, clip]{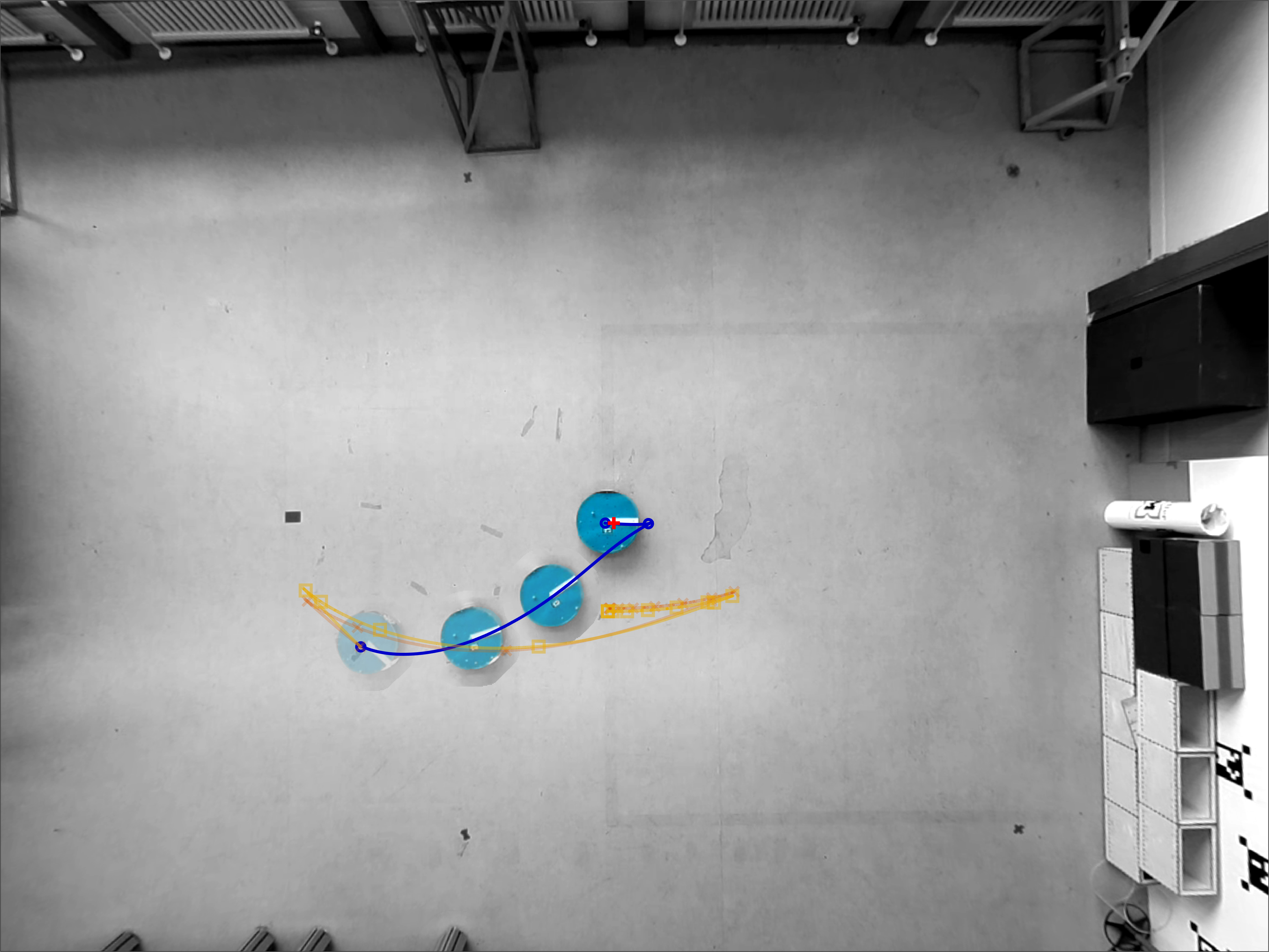}
    \end{minipage}
    \begin{minipage}{0.64\linewidth}
		\ifbool{compilePlots}{ 
			\tikzsetnextfilename{lab_dynamic}
			\include{figures/lab_dynamic}
		}{
			\includegraphics{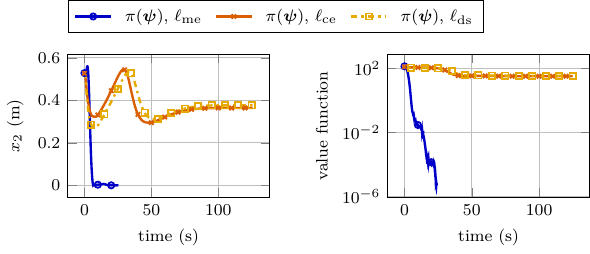}
		}
    \end{minipage}
    \caption{Hardware results. Experimental closed-loop results for parallel parking the second-order dynamics mobile robot with different predictive controllers. Left: Scene stills and planar plot of the EDMD-based controllers. Middle: Deviation in the hard-to-control $x_2$-direction. Right: Value function of the underlying OCPs.}
    \label{fig:dynamic_lab}
\end{figure*}
Lastly, the following experimental parking scenario demonstrates that the previous considerations also fundamentally hold when controlling a real-world second-order dynamics robot.
Therein, the different predictive controllers shall park the mobile robot from an arbitrarily chosen initial configuration $\bm{z}_0 \coloneqq \begin{bmatrix} \unit[-1.1]{m} & \unit[-0.53]{m} & \unit[-42.68]{deg} & \unitfrac[0]{m}{s} & \unitfrac[0]{rad}{s}\end{bmatrix}^\top$ to the origin, where for all runs the robot was initially placed by hand within an accuracy of $\unit[7]{mm}$ and $\unit[5]{deg}$.
In order to have the full state available for the underlying optimal control problem~\eqref{eq:OCP}, in addition to the robot's pose measured by the external tracking system, the robot's onboard program publishes the wheels' angular velocity based on the motors' encoder values averaged over $\unit[50]{ms}$ such that its velocities $v$ and $\omega$ follow from the kinematic relation~\eqref{eq:kinematic_relation}.
Figure~\ref{fig:dynamic_lab} shows that only predictive controllers subject to the tailored mixed-exponents cost~\eqref{eq:costME_dynamic} are capable of reliably driving the second-order dynamics mobile robot to the desired setpoint.
Controllers subject to a quadratic error metric fail to stabilize the origin.
In particular, the resulting closed-loop behavior of these controllers is considerably worse than for the analogous controllers of the kinematic mobile robot, cf. Section~\ref{sec:results_kinematic}.
As Figure~\ref{fig:dynamic_lab} indicates, these controllers yield a remaining deviation of not only several centimeters in the hard-to-control $x_2$-direction, but rather not even drive the robot close to the origin. 
Again, the illustration on the left shows the overlay of experiment still images for the resulting closed-loop trajectory of the EDMD-based predictive controller subject to the mixed-exponents cost, where the robot is depicted with an increasing opacity over time and the desired setpoint is indicated by the red cross plotted on top.
Additionally, the closed-loop trajectories of the EDMD-based controllers subject to the quadratic costs~$\ell_{\textnormal{ce}}$ and $\ell_{\textnormal{ds}}$ are depicted, clearly showing no convergence to the origin. 

\section{Data-Efficiency of the Control-Affine EDMD Approach}\label{sec:data_efficiency}
\begin{figure*}
    \centering
    \ifbool{compilePlots}{ 
		\tikzsetnextfilename{data_efficiency}
        \include{figures/ecdf_data_efficiency}
    }{
        \includegraphics{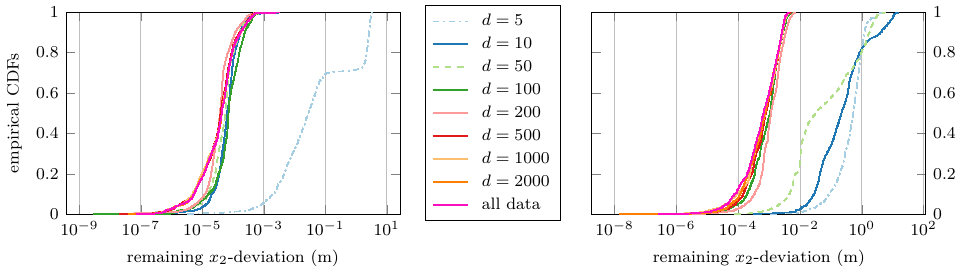}
    }
    \caption{Simulation. Empirical cumulative distribution functions of the deviation in the hard-to-control $x_2$-direction when parking the nonholonomic robots to the origin using EDMD-based predictive controllers with surrogate models identified by having different number of data points available. 
    Left: Deviation after $\unit[10]{s}$ when parking the kinematic mobile robot~\eqref{eq:kinematic_model}. 
    Right: Deviation after $\unit[20]{s}$ when parking the dynamic mobile robot~\eqref{eq:2order_model}.}
    \label{fig:CL_boxplot_data_efficiency}
\end{figure*}
One of the main advantages of the EDMD approach for control-affine systems described in Section~\ref{sec:EDMD} is 
that the curse of dimensionality is moderated by not sampling over the whole input space but rather learning $n_u+1$ autonomous systems using the input bases~$\bm{u}_i$, $i\in\mathbb{Z}_{0:n_u}$.
In combination with the straightforward solution of~\eqref{eq:propagation_Koopman}, this renders the present scheme very data-efficient and computationally inexpensive compared to other learning or identification approaches.
The data-efficiency is elucidated by the same consideration as previously, i.e., considering the remaining deviation in the hard-to-control $x_2$-direction, but now for surrogate models within the predictive controllers that have been learned for different amounts of training data.
For 1000 randomly (i.i.d.) drawn initial poses $\bm{x}_0\in\mathcal{X}_0$ and $\bm{z}_0\in\mathcal{Z}_0$, EDMD-based predictive controllers shall drive the kinematic and the second-order dynamics robot to the origin.
Both controllers utilize the tailored mixed-exponents costs~\eqref{eq:costME} within their respective OCP, reprojection is done in each step, the control parameters are the same as in Sections~\ref{sec:results_kinematic} and~\ref{sec:results_dynamic}, and the respective dictionaries $\mathcal{D}_{5,\textnormal{t}}$ and $\mathcal{D}_\textnormal{Eul}$ are used.
Crucially, the controllers' surrogate models are learned with a different number of data points, where the labels given in Figure~\ref{fig:CL_boxplot_data_efficiency} describe the number $d$ of data points available for each of the $n_u+1$ autonomous subsystems, see also Section~\ref{sec:EDMD}.
As can be seen, already for a very small number of data points, the EDMD-based predictive controllers reliably steer the robots (close) to the origin.
For the kinematic mobile robot, already ten measurements per autonomous subsystem, which, at best, can be achieved by taking measurements over one second, are sufficient to identify a precise model, see Figure~\ref{fig:CL_boxplot_data_efficiency} (left).
Analogously, for the second-order dynamics robot, taking 100 data points (which is equivalent to a time span of $\unit[5]{s}$) per autonomous subsystem yields already a precise model, see Figure~\ref{fig:CL_boxplot_data_efficiency} (right).

\section{Discussion and Outlook}\label{sec:summary}
We have shown that, by using an EDMD approach tailored to control-affine systems, precise models for differential-drive robots can be identified, not only for kinematic mobile robots but also for second-order dynamics vehicles.
As the paper shows, identifying a second-order dynamics mobile robot requires a more sophisticated approach 
than kinematically approximating a robot. 
This is due to the increased dimensionality, but also since data on velocity-level is needed, which is much harder to sample, measure, and process in the physical world than pure position data. 
However, our 
results have shown that, with appropriate post-processing of the data, useful models can be obtained. 
As it turned out, as we use real-world, noisy data, the obtained open-loop predictions are highly sensitive w.r.t.\  
some of the post-processing's hyperparameters, although, in the end, it was easy to pick reliably working ones. 
The effects of some of these hyperparameters, e.g., the smoothing of the approximated velocities, as well as the quality of different dictionaries for the second-order dynamics mobile robot have  been investigated extensively by means of the resulting open-loop predictions for two different real-world reference trajectories.
In addition to the one-step errors, also multi-step errors have been considered in order to examine the possibilty to use the data-driven surrogate models within a predictive controller.

As the main contribution, it has been shown in simulations and hardware experiments that the surrogate models can be utilized to design functioning predictive controllers both for a kinematically treated mobile robot and for a second-order dynamics treatment that can account structure-wise for actuator dynamics. 

Still, the results seriously draw into question a purely data-centric approach that foregoes systematic a-priori knowledge altogether. 
It turned out that, although one can identify a (precise) model purely from data without first-principles insight, knowledge about the underlying geometry is still indispensable to use the learned model to obtain a high-precision predictive controller.
This has been exemplified for both robot types by comparing controllers subject to the conventional quadratic cost choices for the state space and the lifted descriptions, respectively, which clearly fail to reliably drive the vehicles to their goal.
Only controllers subject to a tailored mixed-exponents cost, which is established based on the underlying sub-Riemannian geometry, are able to reliably control the robots -- also for the considered data-driven predictive controllers. 

Furthermore, it has been shown that precise control is possible even for very few data points to identify the prediction model. 
This confirms empirically with real-world hardware and data one of the supposed main advantages of the considered EDMD approach for control-affine systems. 
Therefore, the approach does not only seem very data efficient and practical for nonholonomic robots that are (approximately) input affine, but it also presents categorical advantages compared to some other data-driven MPC approaches like the popular DeePC~\cite{CoulLyge19} or (linear) subspace predictive control~\cite{FiedLuci21}, which structurally, fundamentally rely on linear system approximations and, thus, are by definition troubled with nonholonomic systems for which linearizations are not even locally controllable. 

However, despite these structural advantages, future work with this and other data-driven schemes should focus on robotics and engineering applications that are particularly hard or impossible to solve without predominantly data-driven methods. 
At the very least, this work shows that the proposed methodology is a good candidate for such challenges, and having studied scenarios still allowing baseline comparisons will be a good foundation. 
Moreover, formations of nonholonomic robot may be be considered~\cite{RoseEbel22b,EbelRose23b} in a data-driven manner.
Herein, symmetry exploitation may be of key importance as already demonstrated for, e.g., translational invariance, see also~\cite{hochrainer2024approximation}. The respective physical insights on symmetry may then be leveraged in the construction of suitable dictionaries and w.r.t.\ data requirements, see, e.g., 
\cite{salova2019koopman,peitz2023partial}.

\section{Acknowledgments}
The ITM acknowledges the support by the Deutsche Forschungsgemeinschaft (DFG, German Research Foundation) under Germany’s Excellence Strategy – EXC 2075 – 390740016, project PN4-4 “Learning from Data - Predictive Control in Adaptive Multi-Agent Scenarios” as well as project EB195/32-1, 433183605 “Research on Multibody Dynamics and Control for Collaborative Elastic Object Transportation by a Heterogeneous Swarm with Aerial and Land-Based Mobile Robots” and project EB195/40-1, 501890093 “Mehr Intelligenz wagen - Designassistenten in Mechanik und Dynamik (SPP 2353)”. K.\ Worthmann gratefully acknowledges funding by the German Research Foundation (DFG, project-ID 507037103).

\end{document}